# How disinformation and fake news impact public policies?: A review of international literature [English]

*Ergon Cugler de Moraes Silva*

Getulio Vargas Foundation (FGV)
University of São Paulo (USP)
São Paulo, São Paulo, Brazil

contato@ergoncugler.com
www.ergoncugler.com

*José Carlos Vaz*

University of São Paulo (USP)
Getulio Vargas Foundation (FGV)
São Paulo, São Paulo, Brazil

vaz@usp.br
www.getip.net.br

**Abstract**

This study investigates the impact of disinformation on public policies. Using 28 sets of keywords in eight databases, a systematic review was carried out following the "Prisma 2020" model (Page et al., 2021). After applying filters and inclusion and exclusion criteria to 4,128 articles and materials found, 46 publications were analyzed, resulting in 23 disinformation impact categories. These categories were organized into two main axes: "State and Society" and "Actors and Dynamics", covering impacts on State actors, society actors, State dynamics and society dynamics. The results indicate that disinformation affects public decisions, adherence to policies, prestige of institutions, perception of reality, consumption, public health and other aspects. Furthermore, this study suggests that disinformation should be treated as a public problem and incorporated into the public policy research agenda, contributing to the development of strategies to mitigate its effects on government actions.

## 1. Introduction

This study aims to investigate how disinformation affects public policies. For this, 28 sets of keywords were selected, applied to eight different databases, resulting in a systematic review based on the "Prisma 2020" model (Page et al., 2021). After two rounds of applying filters and strict inclusion and exclusion criteria, 46 relevant publications were identified. These publications were grouped by themes and analyzed, culminating in the definition of 23 analytical categories to observe the impacts of disinformation on public policies.

To deepen the analysis, these 23 categories were organized into two main axes: "State" and "Society", in addition to "Actors" and "Dynamics". This framework allowed a detailed observation of the impacts of disinformation in four distinct areas, these being: **a) impacts on State actors; b) impacts on societal actors; c) impacts on State dynamics; and d) impacts on societal dynamics.**

**Impacts on State actors:** It has been observed that disinformation can significantly impact State actors, including judiciary agents, regulators and bureaucrats at different levels. For example, judicial agents can make decisions based on false information, compromising access to justice and the credibility of institutions. One notable case involves court decisions influenced by conspiracy theories or fake news, which can result in erroneous sentences and a loss of public trust in the judicial system. Furthermore, regulatory instances, by basing their policies on incorrect data, may implement inappropriate regulations that affect critical sectors of the economy and public safety.

**Impacts on actors in society:** It has also been observed that disinformation affects important actors in society, such as the media, civil society organizations and the private sector. The media, for example, can be affected by false information, influencing public opinion and directing public debate towards misleading topics. A narrative dispute became evident during the COVID-19 pandemic, for example, where the dissemination of false information about treatments and vaccines affected the population's adherence to public health measures. Civil society organizations can be misled by false data, compromising their campaigns and advocacy efforts. Private companies, when making decisions based on disinformation, can face financial losses and contribute to economic destabilization.

**Impacts on State dynamics:** In State dynamics, disinformation can affect institutional communication and public policy planning. Institutional communication is crucial to ensure that correct information reaches the population. When disinformation prevails, it is observed that the State needs to invest more in clarification campaigns, such as those observed during vaccination campaigns, where it was necessary to combat myths and disinformation. This represents a diversion of resources that could be used in other essential areas. Furthermore, disinformation can compromise public planning, leading to misallocation of resources and ineffective implementation of policies, as evidenced in public health crises where the state response has been hampered by incorrect information.

**Impacts on societal dynamics:** Disinformation can also significantly alter the perception of reality and the population's adherence to public policies. For example, false information about vaccine safety can reduce the rate of vaccination, resulting in the resumption of diseases that have already been overcome. This has been seen with the resurgence of diseases such as measles in regions where anti-vaccine movements have gained strength. Furthermore, disinformation can polarize public opinion, reinforcing incorrect beliefs and hindering consensus on important political and social issues. This polarization can lead to social conflicts and hinder the implementation of evidence-based public policies, harming collective well-being and social cohesion.

In this way, this study suggests that disinformation should be recognized as a significant public problem and incorporated into the public policy research agenda. Understanding and mitigating the impacts of disinformation can contribute to the development of more effective strategies in the formulation and implementation of public policies, thus strengthening governance and trust in public institutions.

## 2. Materials and methods

Based on the "integrative systematic review" proposed by Botelho et. al (2011, p. 127), a search was carried out in titles and abstracts of publications, without temporal restrictions, using 28 searches with the Boolean operator 'AND'. The Figure 01 details.

**Figure 01.** "Prisma 2020 Framework" model for systematic literature review:

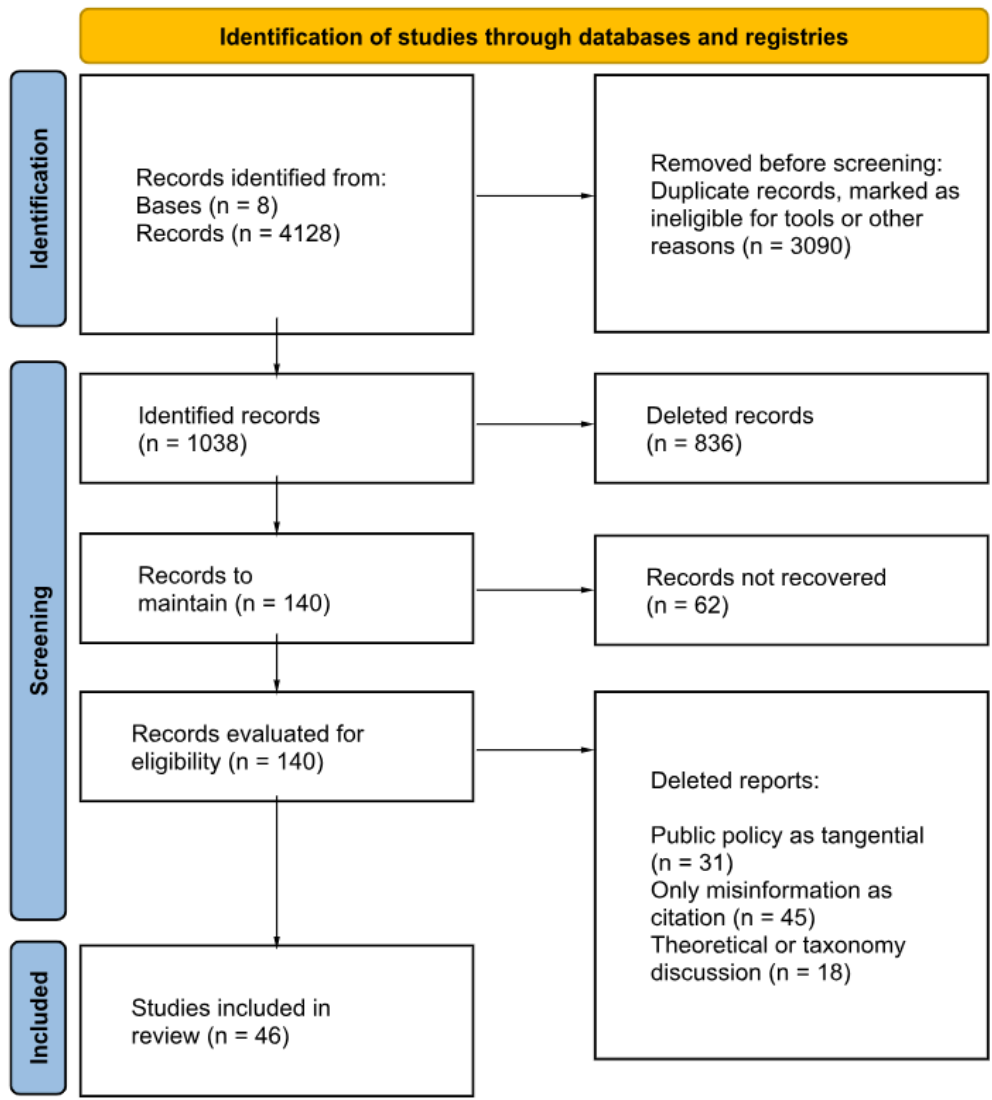

Source: Authors (2024).

The keywords were listed, being in **portuguese:** "fake news" + "literatura"; "desinformação" + "literatura"; "agnotologia" + "literatura"; "fake news" + "efeito"; "desinformação" + "efeito"; "agnotologia" + "efeito"; "fake news" + "impacto"; "desinformação" + "impacto"; "agnotologia" + "impacto"; "fake news" + "políticas públicas"; "desinformação" + "políticas públicas"; "agnotologia" + "políticas públicas". In **english:** "fake news" + "literature review"; "misinformation" + "literature review"; "disinformation" + "literature review"; "agnotology" + "literature review"; "fake news" + "effect"; "misinformation" + "effect"; "disinformation" + "effect"; "agnotology" + "effect"; "fake news" + "impact"; "misinformation" + "impact"; "disinformation" + "impact";

"agnotology" + "impact"; "fake news" + "public policy"; "misinformation" + "public policy"; "disinformation" + "public policy"; "agnotology" + "public policy".

The proposal for the "Prisma 2020" method (Page et al., 2021) was also used, which is a set of evidence-based steps to conduct reports in systematic reviews. These 28 sets of keywords were searched in 8 databases: 1.) Google Scholar; 2.) ProQuest; 3.) Scielo (Scientific Electronic Library Online); 4.) Capes Periodical Portal; 5.) Capes Theses & Dissertations Catalog; 6.) Portal of the Digital Library of Theses and Dissertations of the University of São Paulo (USP); 7.) Oasis BR Ibict; and 8.) Brazilian Digital Library of Theses and Dissertations (BDTD). An initial survey brought a total of 4,128 results with keywords in the title or abstract and, systematically, 46 studies were selected to be reviewed.

## 3. Exploring the results

**Table 01.** Analysis categories:

| Category | Description | Reference used |
|---|---|---|
| Impacts on **Judicial agents** | Impacts on state actors allocated in the judiciary and their decision-making and influence on public policies and government actions (e.g., public decisions based on disinformation; or omissions in the face of disinformation; or positions sustained by belief systems in contrast to scientific knowledge). | **1.** Oriras et al., 2018; **2.** Castinholi, 2019; **3.** Araujo et al., 2020. |
| Impacts on **regulatory and oversight agents** | Impacts on state actors allocated in regulatory and oversight bodies (e.g., state-owned enterprises, agencies, and others) and their decision-making and influence on public policies and government actions (e.g., public decisions based on disinformation; or omissions in the face of disinformation; or positions sustained by belief systems in contrast to scientific knowledge). | **1.** Oriras et al., 2018; **2.** Lewandowsky, 2021; **3.** Soares, 2020; **4.** Vignoli et al., 2021; **5.** Castinholi, 2019. |
| Impacts on **mid-level bureaucrats** | Impacts on state actors allocated in mid-level bureaucracy (e.g., department managers and others) and their decision-making and influence on public policies and government actions (e.g., public decisions based on disinformation; or omissions in the face of disinformation; or positions sustained by belief systems in contrast to scientific knowledge). | **1.** Rodríguez-Fernández, 2019; **2.** Soares, 2020; **3.** Keenan et al., 2018; **4.** Vignoli et al., 2021. |
| Impacts on **street-level bureaucrats** | Impacts on state actors allocated in street-level bureaucracy (e.g., police officers, | **1.** Keenan; Dillenburger, 2018. |

| Category | Description | Reference used |
|---|---|---|
| | teachers, healthcare professionals, and others) and their decision-making and influence on public policies and government actions (e.g., public decisions based on disinformation; or omissions in the face of disinformation; or positions sustained by belief systems in contrast to scientific knowledge). | |
| Impacts on **policymakers bureaucrats** | Impacts on state actors allocated in high-level bureaucracy (e.g., secretaries, ministers, directors of state-owned enterprises, policymakers involved in the design of public policies) and their decision-making and influence on public policies and government actions (e.g., public decisions based on disinformation; or omissions in the face of disinformation; or positions sustained by belief systems in contrast to scientific knowledge). | **1.** Rodríguez-Fernández, 2019; **2.** Soares, 2020; **3.** Vignoli et al., 2021; **4.** Castinholi, 2019; **5.** Araujo et al., 2020. |
| Impacts on **institutional councils and committees** | Impacts on state actors and society actors composing elective committees and institutional councils and their decision-making and influence on public policies and government actions (e.g., public decisions based on disinformation; or omissions in the face of disinformation; or positions sustained by belief systems in contrast to scientific knowledge). | **1.** Rodríguez-Fernández, 2019; **2.** Soares, 2020; **3.** Vignoli et al., 2021, **4.** Araujo et al., 2020. |
| Impacts on **legislators** | Impacts on state actors with elective and representative functions in the legislature and their decision-making and influence on public policies and government actions (e.g., public decisions based on disinformation; or omissions in the face of disinformation; or positions sustained by belief systems in contrast to scientific knowledge). | **1.** Lewandowsky, 2021; **2.** Vignoli et al., 2021; **3.** Araujo et al., 2020. |
| Impacts on **media and press** | Impacts on the media and press and their influence on public policies and government actions (e.g., positions of sectors on agendas and public policies; mobilization of journalistic agendas around a proposal for approval). | **1.** Pate et al., 2019; **2.** Lisboa et al., 2020; **3.** Mello, 2020. |
| Impacts on **civil society and organizations** | Impacts on society actors in civil society organizations and their influence on public policies and government actions (e.g., public decisions based on | **1.** Rodríguez-Fernández, 2019; **2.** Vignoli et al., 2021; **3.** Araujo et al., 2020. |

| Category | Description | Reference used |
|---|---|---|
| | omissions in the face of disinformation; or positions sustained by belief systems in contrast to scientific knowledge). | |
| Impacts on **political parties and epistemic communities** | Impacts on state and society actors composing political parties and/or epistemic communities and their influence on public policies and government actions (e.g., public decisions based on disinformation; or omissions in the face of disinformation; or positions sustained by belief systems in contrast to scientific knowledge). | **1.** Rodríguez-Fernández, 2019; **2.** Vignoli et al., 2021, **3.** Araujo et al., 2020. |
| Impacts on **private sector and corporations** | Impacts on the private sector and their influence on public policies and government actions (e.g., corporations' positions on agendas and public policies; mobilization of the business sector around a proposal for approval). | **1.** Rodríguez-Fernández, 2019. |
| Impacts on **institutional communication and ICTs** | Impacts on the state structure concerning institutional communication, public relations, and/or demands for state capacities for practices and Information and Communication Technologies to counter disinformation (e.g., anti-fake news campaigns; or fact-checking technologies). | **1.** Carneiro, 2019; **2.** Silva, 2017; **3.** Vaccari et al., 2020; **4.** Whyte, 2020; **5.** Matasick et al., 2020; **6.** Vignoli et al., 2021. |
| Impacts on **public budget and expenses** | Impacts on the state structure concerning public budget and expenses (e.g., costs of activities to combat disinformation; or loss of assets leading to state costs). | **1.** Vaccari et al., 2020; **2.** Whyte, 2020; **3.** Matasick et al., 2020. |
| Impacts on **public planning and resources** | Impacts on the state structure concerning public planning (e.g., goals, predictability, crisis management) and allocation of resources and inputs (e.g., human resources, materials, or state demands). | **1.** Matasick et al., 2020. |
| Impacts on **prestige of Institutions and the System** | Impacts on the state structure concerning the prestige of institutions and democracy (e.g., hate speech; institutional fragility; state demobilization; or attacks on democracy). | **1.** Rodríguez-Fernández, 2019; **2.** Milla et al., 2020; **3.** Pate et al., 2019; **4.** Silva, 2019; **5.** Patihis et al., 2018; **6.** Benedict et al., 2019; **7.** Greenspan et al., 2020; **8.** Prandi et al., 2020; **9.** Ferreira et al., 2021; **10.** Than et al., 2020; **11.** Ghenai et al., 2017; **12.** McNamara, 2019; **13.** Miller, 2019; **14.** Newton, 2019; **15.** Huang et al., 2019; **16.** Souza Junior et al., 2020; **17.** Souza, |

| Category | Description | Reference used |
|---|---|---|
| | | 2020; **18.** Patel., 2020; **19.** Araujo et al., 2020; **20.** Da Empoli, 2019; **21.** Norris et al., 2019; **22.** Benkler et al., 2018; **23.** Alves, 2020. |
| Impacts on **domestic and/or foreign relations** | Impacts on the state structure in relationships between entities (e.g., between spheres; between branches; between external entities; between peoples; or global multilateral organizations). | **1.** Cadwalladr et al., 2018; **2.** Valente, 2019; **3.** Pate et al., 2019; **4.** Mejias et al., 2017; **5.** La Cour, 2020; **6.** Landon-Murray et al., 2019. |
| Impacts on **adherence to public policies** | Impacts on the structure of society concerning citizens' adherence to public policies (e.g., adherence to vaccination campaigns; or state guidelines; alignment with national agendas). | **1.** Keenan et al., 2018; 2. Pate et al., 2019; **3.** Prandi et al., 2020; **4.** Silva et al., 2017; **5.** Guimarães, 2017, apud Silva et al., 2017; **6.** Vignoli et al., 2021; **7.** Than et al., 2020; **8.** Ghenai et al., 2017; **9.** McNamara, 2019; **10.** Miller, 2019; **11.** Newton, 2019; **12.** Huang et al., 2019; **13.** Souza Junior et al., 2020; **14.** Souza, 2020, apud Almeida et al., 2020. |
| Impacts on **changes in the production chain** | Impacts on the structure of society concerning production chains, including the relationship of such chains with the logistical production flows of the state (e.g., production chain stoppages/strikes). | **1.** Almeida et al., 2020; **2.** Zhang, 2020, apud Almeida et al., 2020. |
| Impacts on **changes in perception of reality** | Impacts on the structure of society concerning changes, including psychological ones, in the perception of reality (e.g., narratives disputing the truth; conflicts over ambiguous guidelines; or psychological and memory impacts on the perception of reality). | **1.** Silva, 2019; **2.** Patihis et al., 2018; **3.** Benedict et al., 2019; **4.** Greenspan et al., 2020; **5.** Pate et al., 2019; **6.** Almeida et al., 2020; **7.** Peeri et al., 2020; **8.** Santos et al., 2020; **9.** Matos et al., 2020; **10.** Neto et al., 2020; **11.** Sutherland et al., 2001; **12.** Wylie et al., 2014, apud Greenspan et al., 2020; **13.** Vignoli et al., 2021; **14.** Patel., 2020; **15.** Araujo et al., 2020; **16.** Da Empoli, 2019; **17.** Norris et al., 2019; **18.** Benkler et al., 2018; **19.** Alves, 2020. |
| Impacts on **consumption patterns** | Impacts on the structure of society concerning changes in social consumption patterns of goods and services (e.g., consumption of medications without specialist prescription; or asymmetries in the consumption of inputs and food, such as cases of rumors about shortages). | **1.** Almeida et al., 2020; **2.** Zaracostas, 2019, apud Almeida et al., 2020. |
| Impacts on | Impacts on individuals and/or social groups | **1.** Lisboa et al., 2020; **2.** Libório et |

| Category | Description | Reference used |
|---|---|---|
| **externalities, integrity, and environment** | concerning their health, integrity, and/or safety (e.g., fatalities or illnesses resulting from disinformation acts; externalities to the environment, social and/or economic; or sectoral externalities related to the observed public policy). | al., 2020, apud Lisboa et al., 2020; **3.** Islam et al., 2020; **4.** Valente, 2019; **5.** Pate et al., 2019; **6.** Mejias et al., 2017; **7.** Almeida et al., 2020; **8.** Peeri et al., 2020; **9.** Santos et al., 2020; **10.** Matos et al., 2020; **11.** Neto et al., 2020; **12.** Zaracostas, 2019; **13.** Than et al., 2020; **14.** Ghenai et al., 2017; **15.** McNamara, 2019; **16.** Miller, 2019; **17.** Newton, 2019; **18.** Huang et al., 2019; **19.** Souza Junior et al., 2020; **20.** Souza, 2020, apud Almeida et al., 2020. |
| Impacts on **public opinion and belief systems** | Impacts on the structure of society concerning the alteration and construction of public opinion and the social imagination amid disputed belief systems (e.g., discrimination against specific peoples and/or social groups; or categorical narratives about a program and public policy; or the perception and public opinion about a particular public policy). | **1.** Oriras et al., 2018; **2.** Milla et al., 2020; **3.** Pate et al., 2019; **4.** Sutherland et al., 2001; **5.** Wylie et al., 2014, apud Greenspan et al., 2020; **6.** Vignoli et al., 2021; **7.** Ferreira et al., 2021; **8.** Castinholi, 2019; **9.** Patel, 2020; **10.** Araujo et al., 2020; **11.** Da Empoli, 2019; **12.** Norris et al., 2019; **13.** Benkler et al., 2018; **14.** Alves, 2020. |
| Impacts on **prestige of scientific knowledge** | Impacts on the structure of society concerning the prestige of science, including the scientific community in its various fields (e.g., conspiracy theories; ambiguous narratives devaluing science; scientifically delegitimized dismantling by rhetoric). | **1.** Almeida et al., 2020; **2.** Peeri et al., 2020; **3.** Santos et al., 2020; **4.** Matos et al., 2020; **5.** Neto et al., 2020; **6.** Sutherland et al., 2001; **7.** Wylie et al., 2014, apud Greenspan et al., 2020; **8.** Silva et al., 2017; **9,** Vignoli et al., 2021; **10.** Patel., 2020; **11.** Araujo et al., 2020; **12.** Da Empoli, 2019; **13.** Norris et al., 2019; **14.** Benkler et al., 2018; **15.** Alves, 2020. |

Source: Own elaboration. *From authors reviewed in their publication (2024).

Furthermore, Figure 02 below illustrates the 23 potential impacts of disinformation:

**Figure 02.** Disinformation impact matrix:

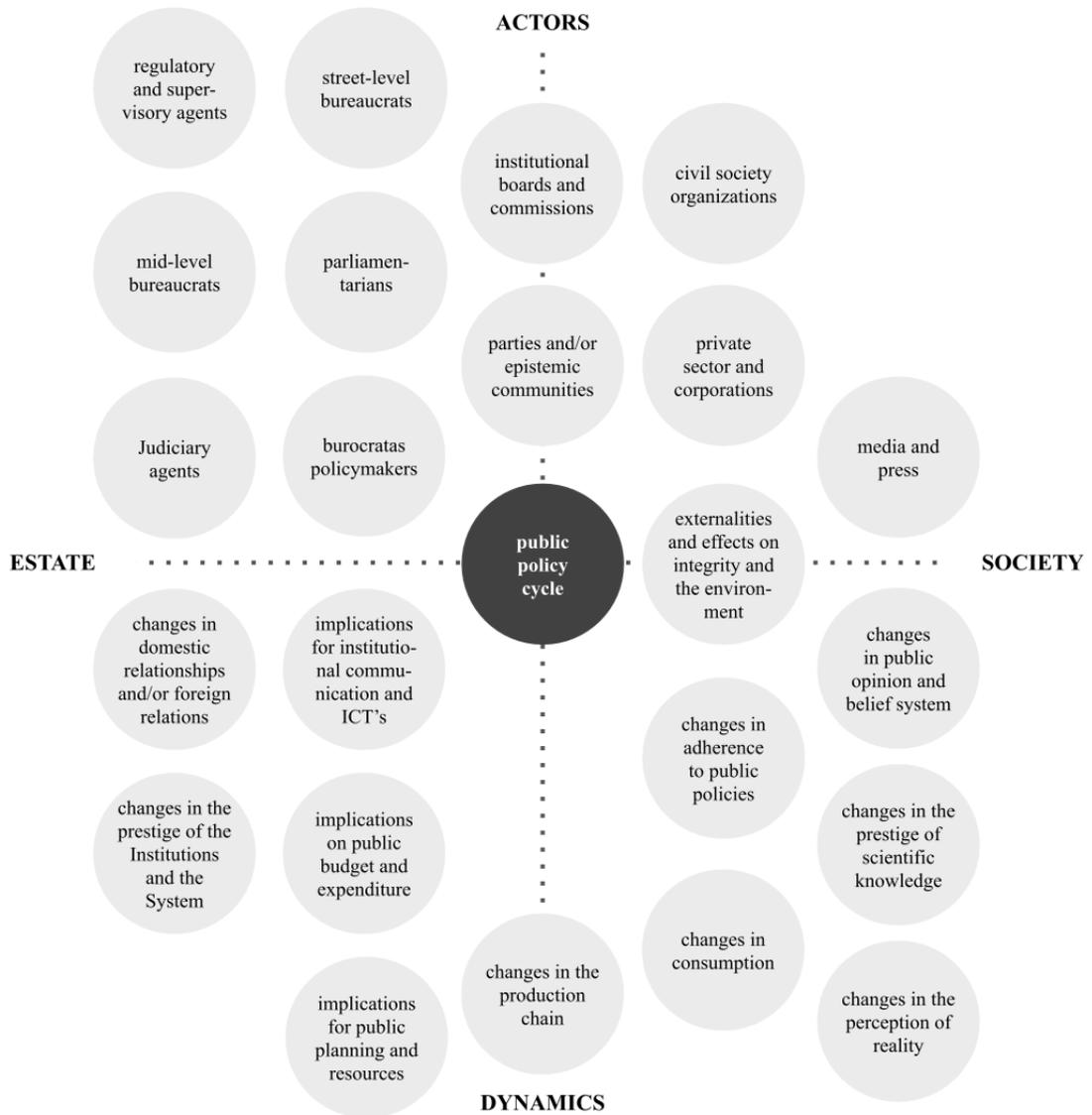

Source: Own elaboration (2024).

## 4. Experimentation with the analytical model

**Table 02.** Test with case of irregular early treatment (off label) in the COVID-19 Pandemic:

| Impacts on State Actors | |
|---|---|
| **Category** | **Observed impacts** |
| Impacts on **Judicial agents** | Discussions about Hydroxychloroquine/Chloroquine have caused turmoil among jurists. The Federal Public Defender's Office (DPU) filed a lawsuit against the |
| | Brazilian Federal Council of Medicine (CFM) for its omission concerning the alleged early treatment. Several critical sectors publicly attacked the DPU's stance [https://revistaoeste.com/politica/mais-uma-aberracao-do-sistema-judiciario/], while the National Council of Justice (CNJ) had to issue a statement to guide judges [https://www.conjur.com.br/2020-mar-23/cnj-divulga-parecer-orientar-juizes-hidroxicloroquina] amidst ambiguities. So, disinformation about early treatment required significant efforts from the Judiciary and resulted in local decisions that legitimized off label based on medical autonomy [https://www.conjur.com.br/2020-mai-30/responsabilizacao-juridica-uso-cloroquina-nao-consenso]. Consequently, disinformation and decisions misaligned with the scientific community's prevailing understanding fragmented the policy for combating the COVID-19 pandemic. Courts relied on their respective judicial decisions until the Supreme Federal Court (STF) officially took a position [https://www.cnnbrasil.com.br/politica/stf-determina-que-governo-bolsonaro-apresente-protocolo-para-tratamento-da-covid/], seeking to address the disinformation that was creating ambiguities among judges and their decisions. This highlights the impact on health policy implementation, dispersing guidelines and norms |
| Impacts on **regulatory and oversight agents** | No impacts from disinformation regarding 'early treatment' on agents and regulatory bodies were found. On the contrary, right at the beginning of the Pandemic (April 2020), the National Health Surveillance Agency (ANVISA) had pointed out to the Bolsonaro Government that Hydroxy/Chloroquine should not be used against COVID-19 [https://www.brasildefato.com.br/2020/08/24/em-abril-anvisa-alertou-bolsonaro-que-cloroquina-nao-deveria-ser-usada-contra-covid]. |
| Impacts on **mid-level bureaucrats** | It is worth highlighting a case involving mid-level bureaucrats from the Ministry of Health (MS), in which such bureaucrats, regardless of their motivations (which are not part of this discussion), but being involved in the "early treatment" agenda, began to obstruct the investigations of the COVID-19 Parliamentary Inquiry Commission (CPI) in the Senate. In response, the CPI voted to remove those employees [https://oglobo.globo.com/politica/cpi-da-covid-votara-afastamento-de-servidores-que-estao-obstruindo-investigacoes-ou-poderiam-obstruir-1-25131794]. It is worth mentioning that the MS has several strategic positions allocated by actors involved in the "early treatment" agenda. In leadership positions, Hélio Angotti and Mayra Pinheiro actively campaign even after almost two years of the Pandemic, in defense of the "Covid Kit" |
| | [https://www.correiobraziliense.com.br/politica/2021/02/4904951-mais-de-20-militares----capita-cloroquina----e-medico-olavista-ocupam-cargos-no-ministerio-da-saude.html] and thus, consequently, they carry an agenda without any scientific evidence for the MS routine, directly affecting public policies. Given the context of circulation of mid-level actors, it is observed how they can impact all stages of the public policy cycle when observing the dispute over "early treatment", since from the formulation of "TrateCov", to its implementation and attempt of distorting the assessment during the investigation, such bureaucrats exerted influence. |
| Impacts on **street-level bureaucrats** | Even a year after the start of the COVID-19 Pandemic, a survey carried out by the Brazilian Medical Association (AMB) showed that among doctors working on the front line, 37.4% believe that chloroquine is effective in preventing Covid-19, while 41.4% see the same potential in ivermectin [https://olhardigital.com.br/2021/02/03/videos/covid-19-1-3-dos-medicos-acredita-em-tratamento-precoce/] even without any scientific evidence to do so. In a survey coordinated by professor Gabriela Lotta (FGV), in which 1,520 Community Health Agents (CHAs) are interviewed, despite the fact that "more than 90% of those interviewed disapprove, totally or partially, of President Jair Bolsonaro prescribing medications for the population", it is observed that a third (33.8%) of the CHWs interviewed do not think that only drugs with proven efficacy should be used, opening a gap for discretion amid discussions about "early treatment" [https://www.revistaquestaodeciencia.com.br/artigo/2020/11/11/pesquisa-da-fgv-mostra-falta-de-treinamento-de-testes-e-medo-entre-profissionais-de-saude-no-sus]. It is worth highlighting not only the impacts on the implementation of public policy, but also the vulnerabilities that inequalities bring in the face of information asynchronies and driven distortions. Communities dependent on SUS capillarity, for example, rely on the work of CHAs to practice their relationship with the State. |
| Impacts on **policymakers bureaucrats** | Among the most evident conflicts and ambiguities in relation to early treatment, the Federal Government's own high-ranking bureaucrats stand out in the internal dispute and with other spheres and Powers of the country. Internally, two Health Ministers were fired for not publicly supporting the supposed 'early treatment' with Hydroxy/Chloroquine [https://www1.folha.uol.com.br/cotidiano/2020/12/tres-ministros-da-saude-e-uma-pandemia-o-ano-em-que-ficamos-doentes.shtml], despite the lack of any evidence or even international scientific organization supporting such a thesis. In view of all the documentation that highlights Federal omission and lack of coordination |
|  | [https://www.camara.leg.br/noticias/766207-instituicoes-sociais-denunciam-omissao-do-governo-no-combate-a-pandemia], There is no doubt about how disinformation guided high-ranking political actors, directly affecting proposals such as that of the Ministry of Health (MS) regarding the "Covid Kit", based on the false thesis of "early treatment" and "herd immunity". It was only months later that the MS itself admitted the ineffectiveness of its own "Covid Kit" [https://pebmed.com.br/ministerio-da-saude-confirma-ineficacia-do-kit-covid-no-tratamento-contra-covid-19/]. In this sense, it is observed how disinformation circulates among formulators, moving from this stage of the public policy cycle to the praxis of implementation. |
| Impacts on **institutional councils and committees** | Considering Institutional Councils and Commissions with an impact on public policies related to the topic of Health and the management of the COVID-19 Pandemic, no impacts of disinformation on these were found. On the contrary, the National Health Council (CNS) even filed documents in complaint with the Parliamentary Commission of Inquiry (CPI) on COVID-19 in the Senate along with other organizations and entities [https://www.camara.leg.br/noticias/766207-instituicoes-sociais-denunciam-omissao-do-governo-no-combate-a-pandemia]. Furthermore, it was the CNS itself that "asked [for] the Ministry of Health to revoke the technical note that guided the use of chloroquine and to refrain from encouraging the use of medicines without proven efficacy and safety" [https://www.nexojornal.com.br/expresso/2021/05/24/Qual-a-cronologia-cient%C3%ADfica-da-cloroquina-na-pandemia]. Amidst the dispute over the formulation of health policy guidelines, this stage of the public policy cycle is also vulnerable. |
| Impacts on **legislators** | Federal Parliamentarians from the Government's support base, in line with the thesis of "herd immunity" and "early treatment", even presented a Bill to authorize the distribution of Hydroxy/Chloroquine [https://www25.senado.leg.br/web/atividade/materias/-/materia/141533], the discussions of the COVID-19 Parliamentary Commission of Inquiry (CPI) in the Senate left no doubt about how Hydroxy/Chloroquine was adopted as a platform by parliamentarians supporting the Bolsonaro Government, even leading the president of the CPI to present a Bill in the antithesis: prohibiting and criminalizing the encouragement of the use of medicines without scientific evidence [https://oglobo.globo.com/brasil/presidente-da-cpi-apresenta-projeto-que-criminaliza-incentivo-ao-uso-de-remedios-sem-comprovacao-cientifica-25028819]. In short, in addition to the conflicts and ambiguities generated by disinformation about "early treatment" - without any scientific evidence to |

date in relation to COVID-19 - and the existence of parliamentarians based on such theses to design public policies, Parliament's vulnerability to such rhetorical disputes on this issue is observed.

**Impacts on Society Actors**

| Category | Observed impacts |
|---|---|
| Impacts on **media and press** | With the exception of ideologically positioned alternative media with practices of disseminating disinformation, it was rare to observe institutional and public stances in defense of early treatment in the traditional media. Even so, professor Carlos Eduardo Lins da Silva (USP) highlights that the Brazilian media has made the same mistake as the American media and given "a lot of space and prominence to comment on denialists, flat earthers or those in favor of the use of chloroquine. It is necessary to carefully weigh who the defenders of one theory or another are. 'Not giving the same space in the media to those who approve the use of the medicine to the detriment of serious scientific entities'" [https://jornal.usp.br/radio-usp/colunista-analisa-imprensa-americana-e-brasileira-na-abordagem-da-cloroquina/]. While it is not about adopting an institutional denialist stance, the impact of disinformation is that it is treated as an "opinion" and "point of view", which "must be listened to". In this way, the space provided by traditional media indirectly boosts rhetoric of dissemination of disinformation. |
| Impacts on **civil society and organizations** | On the other hand, several relevant organizations in the area took a stand against the "early treatment" thesis, including the Brazilian Medical Association (AMB), the Brazilian Intensive Care Medicine Association (AMIB) and the Infectious Diseases, Immunology, Pulmonology and Physiology societies. [https://www.nexojornal.com.br/expresso/2021/05/24/Qual-a-cronologia-cient%C3%ADfica-da-cloroquina-na-pandemia]. Furthermore, organizations were articulating themselves in the face of the pandemic, as is the case of the then-founded "Association of Victims and Families of Victims of Covid-19 (Avico Brasil)", which proposed criminal representation against the Bolsonaro Government [https://www.camara.leg.br/noticias/766207-instituicoes-sociais-denunciam-omissao-do-governo-no-combate-a-pandemia]. |
| Impacts on **political parties and epistemic communities** | With regard to epistemic communities, some individuals in the medical community were influenced by an article published in the renowned The Lancet, which defended the use of Hydroxy/Chloroquine. A single article among hundreds of others refuting the drug's effectiveness was |
| | enough to support the selective opinion of those who were looking for some reference to legitimize their own truths. Three days later, The Lancet suspended the research and refuted the published efficacy, retracting [https://www.nexojornal.com.br/expresso/2021/05/24/Qual-a-cronologia-cient%C3%ADfica-da-cloroquina-na-pandemia]. Figures such as Mayra Pinheiro and Nise Yamaguchi gained relevance in the midst of health decisions, especially Pinheiro, as she is the Secretary of Health of the Ministry of Health (MS) and an active campaigner for "early treatment". In addition to Yamaguchi, who was considered to be Minister of Health, and would have suggested changing the Chloroquine leaflet to try to legitimize its use against COVID-19 [https://www.bbc.com/portuguese/brasil-57124296]. Agência Lupa even exposed groups of doctors who were spreading and even sponsoring "false information to defend ineffective treatment against Covid-19" [https://piaui.folha.uol.com.br/lupa/2021/02/23/anuncio-medicos-pela-vida-covid-19/]. In relation to political parties, no major mobilization or prominent position beyond expected divergences. |
| Impacts on **private sector and corporations** | Politically active private sector actors mobilized to raise resources and donations for the "Covid Kit" in support of the Federal Government. An emblematic case is that of Luciano Hang, owner of Rede Havan, who admitted that he had mobilized donations for "early treatment", even though this did not have any scientific evidence [https://www.nexojornal.com.br/expresso/2021/09/29/O-papel-de-Luciano-Hang-na-difus%C3%A3o-de-rem%C3%A9dios-ineficazes]. In this sense, groups of businesspeople sought to influence the management of the COVID-19 Pandemic by embarking on disinformation about the Hydroxy/Chloroquine thesis. |

**Impacts on the State Structure**

| Category | Observed impacts |
|---|---|
| Impacts on **institutional communication and ICTs** | Faced with the contradictions of institutionally supporting "early treatment" without any scientific evidence, the Ministry of Health (MS) after being publicly contradicted for publishing links recommending the use of Hydroxy/Chloroquine simply deleted the publication [https://www1.folha.uol.com.br/equilibrioesaude/2021/05/apos-reportagem-ministerio-da-saude-apaga-links-com-prescricao-de-cloroquina-para-covid.shtml]. Interestingly, in the case observed, it is not disinformation that "mess up" Institutional Communication and makes it have to respond to society; but the opposite, it is the disinformation held by |
| Impacts on **public budget and expenses** | the MS that dismantles Institutional Communication by trivializing such channels with content without scientific validity. In this case, disinformation affects Institutional Communication, as it is vulgarized from the inside out in favor of anti-scientific agendas.<br><br>There was state mobilization and tax exemptions based on the Chloroquine narrative. "[The import tax on chloroquine and azithromycin was zeroed, [with] the initial distribution of 3.4 million units of chloroquine and hydroxychloroquine to hospitals and the expansion of chloroquine production by the chemical laboratories of the Armed Forces in Rio de Janeiro. January" [https://www.nexojornal.com.br/expresso/2021/05/24/Qual-a-cronologia-cient%C3%ADfica-da-cloroquina-na-pandemia]. In addition to the evident impact on the drop in revenue, it is necessary to reinforce the mobilization of the Armed Forces for the large-scale production of Chloroquine. "[There were] more than R$1.5 million in contracts without bidding to produce chloroquine or purchase inputs between March and May 2020." [https://apublica.org/2021/03/o-mapa-da-cloroquina-como-governo-bolsonaro-enviou-28-milhoes-de-comprimidos-para-todo-o-brasil/], including being the target of investigation by the Federal Audit Court (TCU). After massive distribution to Brazilian states, 400 thousand units remained in stock and on the way for disposal [https://www.cnnbrasil.com.br/nacional/sem-demanda-nos-estados-400-mil-comprimidos-de-cloroquina-encalham-no-exercito/]. Furthermore, the Federal Government even indicated R$250 million to implement the "Covid Kit" for "early treatment" with distribution to pharmacies [https://www.istoedinheiro.com.br/saude-preve-gastar-r-250-milhoes-para-por-kit-covid-em-farmacias-populares/]. |
| Impacts on **public planning and resources** | With the adoption of the "early treatment" agenda by the Federal Government, all planning related to the recommendations of the World Health Organization (WHO) was ignored, such as the purchase of vaccines and vaccination protocol [https://g1.globo.com/sp/sao-paulo/noticia/2021/01/17/apos-aprovacao-da-anvisa-governo-de-sp-aplica-1a-dose-da-coronavac-antes-do-inicio-do-plano-nacional-de-vacinacao.ghtml], or even social isolation practices and/or mass distribution policies of PPE. In this case, the thesis of "early treatment" exactly drove the omission in planning for the real confrontation of the Pandemic and, consequently, the lack of coordination in the mobilization and distribution of human and material resources. No wonder, there was speed to send Hydroxy/Chloroquine to the states [https://congressoemfoco.uol.com.br/area/governo/documento-detalha-distribuicao-de-cloroquina-a-estados/], while |
| | vaccines were confused between states, damaging the vaccination logistics chain [https://oglobo.globo.com/saude/vacina/ministerio-da-saude-troca-remessa-de-vacinas-manda-doses-do-amazonas-para-amapa-24897507]. One year after the start of the Pandemic, the malaria program ran out of Chloroquine stock after the Government diverted attention to the "Covid Kit" [https://saude.ig.com.br/coronavirus/2021-03-28/programa-de-malaria-ficou-sem-estoque-de-cloroquina-apos-desvio-do-governo.html]. |
| Impacts on **prestige of Institutions and the System** | With the demoralization of the Ministry of Health (MS) after having two Ministers fired for not supporting "early treatment" and the "back and forth" rhetoric [https://www1.folha.uol.com.br/equilibrioesaude/2021/05/apos-reportagem-ministerio-da-saude-apaga-links-com-prescricao-de-cloroquina-para-covid.shtml] Given the daily evidence against "early treatment", the MS's own practices based on disinformation demoralize its actions and the actions of the State. Furthermore, the dispute with the National Health Surveillance Agency (ANVISA) and Bolsonaro's increased tone by publicly disrespecting ANVISA's own health standards almost 20 times [https://noticias.uol.com.br/cotidiano/ultimas-noticias/reporter-brasil/2021/08/04/bolsonaro-violou-norma-da-anvisa-ao-defender-cloroquina-para-covid.htm] it discredits the role of the State and sends the message that anyone can trivialize it. With the demoralization of the then "early treatment" and pressure from the COVID-19 Parliamentary Inquiry Commission (CPI) in the Senate, "chloroquine became a pushing game between [Ministries of] Defense and Health", avoiding responsibility [https://cnts.org.br/noticias/documentos-mostram-que-cloroquina-virou-jogo-de-empurra-entre-defesa-e-saude-apos-pressao-de-cpi/] and accentuating the Federal disarticulation. |
| Impacts on **domestic and/or foreign relations** | The Hydroxy/Chloroquine narrative driven by Donald Trump led Bolsonaro to adopt this thesis "just two days after Trump spoke about chloroquine for the first time" [https://www.nexojornal.com.br/expresso/2021/05/24/Qual-a-cronologia-cient%C3%ADfica-da-cloroquina-na-pandemia]. More than a year later, even with Trump having abandoned the thesis, the Bolsonaro Government continued to persist in the cause. The World Health Organization (WHO) and the European Union reinforced their concerns about Brazil, [https://www.cnnbrasil.com.br/saude/entenda-as-recomendacoes-das-instituicoes-de-saude-contra-o-uso-de-cloroquina/], These concerns add to the others that gradually began to isolate Brazil in foreign relations [https://www.bbc.com/portuguese/internacional-59106310]. |

**Impacts on the Structure of Society**

| Category | Observed impacts |
|---|---|
| Impacts on **adherence to public policies** | As a result of the drive for "early treatment" in the country, the Bolsonaro Government managed to mobilize at least 23% of the population to use medications without scientific evidence, a portion of which admitted to having "used medications for early treatment against COVID-19" [https://g1.globo.com/ciencia-e-saude/noticia/2021/05/19/datafolha-um-em-cada-quatro-brasileiros-usou-remedios-para-tratamento-precoce-contra-a-covid.ghtml]. Interestingly, in the case of "early treatment", it is not only about observing potential non-adherence to public policies and government recommendations, but also observing adherence to a policy based on disinformation, that is, adherence to medications without evidence. scientific. |
| Impacts on **changes in the production chain** | With the support of the Federal Government and insistent interest in the then Hydroxy/Chloroquine [https://politica.estadao.com.br/noticias/geral,documento-de-general-expoe-mapa-da-cloroquina-e-a-cadeia-de-comando-para-produzi-la,70003564204], investigations by the Federal Audit Court (TCU) indicate that specific groups and pharmaceutical companies had moved even before the legal purchase process was opened. In this way, disinformation, supported by institutional rhetoric, induced the sector and resulted in a cost to the State of "167% more than the previous purchase [of Cloroquina]" [https://www.cnnbrasil.com.br/nacional/fornecedora-de-cloroquina-do-exercito-foi-consultada-um-mes-antes-de-concorrente/]. The BNDES even studied the granting of loans to the pharmaceutical sector for the production of Hydroxy/Chloroquine, an action that could also have involved stimulating the production chain [https://aberto.bndes.gov.br/aberto/caso/farmaceuticas/]. It is worth mentioning that internationally relevant sectors in the market, such as the well-known Kodak, showed interest in producing Hydroxy/Chloroquine in light of the institutional interest persisted by then president Donald Trump [https://www.frontliner.com.br/kodak-ira-fabricar-ingredientes-da-cloroquina/] and, even though it is not in Brazil, it reinforces the attention of the productive sector. |
| Impacts on **changes in perception of reality** | The rhetoric of "early treatment" and "herd immunity" added to the other conspiracy theories that circulated the COVID-19 Pandemic: "more than half of Brazilians (56%) believe in the conspiracy theory that hospitals are paid to inflate data on patients who died from COVID-19"; "44.3% of people living in the country believe that there is a 'left-wing conspiracy' that wants to 'take power' in Brazil"; |
| | "50.7% of Brazilians believe that the Chinese government created the coronavirus" [https://www.otempo.com.br/brasil/covid-56-dos-brasileiros-creem-em-teoria-da-conspiracao-que-dados-sao-inflados-1.2483512]. In this sense, the distortion of Brazilians' perception of reality is made clear, as they become accustomed to countless fake news on a daily basis, with difficulties in distinguishing this from reality. |
| Impacts on **consumption patterns** | There was a direct and explicit impact on demand, as it was stimulated through the Federal Government's "Covid Kit", leading to a 358% increase in private consumption of the medicine [https://politica.estadao.com.br/noticias/geral,quem-sao-os-empresarios-que-ganham-com-a-cloroquina-no-brasil,70003360677]. |
| Impacts on **externalities, integrity, and environment** | At the end of 2020, a 558% increase in annual records of adverse effects resulting from Hydroxy/Chloroquine was reported, even causing deaths due to uninformed use. Among the side effects observed are disorders of the nervous, gastrointestinal, psychiatric and cardiac systems [https://www.em.com.br/app/noticia/bem-viver/2021/04/05/interna_bem_viver,1253798/efeitos-adversos-a-cloroquina-disparam-558-e-mortes-sao-registradas.shtml]. |
| Impacts on **public opinion and belief systems** | As this is the consumption of a medicine with no predicted scientific efficacy, the 358% increase in private consumption of the medicine [https://politica.estadao.com.br/noticias/geral,quem-sao-os-empresarios-que-ganham-com-a-cloroquina-no-brasil,70003360677] and the fact that at least 23% of the population consumed this medicine without scientific evidence, given the spread of disinformation [https://g1.globo.com/ciencia-e-saude/noticia/2021/05/19/datafolha-um-em-cada-quatro-brasileiros-usou-remedios-para-tratamento-precoce-contra-a-covid.ghtml]. As pointed out, contexts of greater social vulnerability absorb disinformation with greater incidence [https://www.scielo.br/j/ciedu/a/bW5YKH7YdQ5yZwkJY5LjTts/?lang=pt], reinforcing belief systems especially for this social stratum. |
| Impacts on **prestige of scientific knowledge** | In an explicit case of attack on Science, the infectious disease doctor and research scientist, Marcus Vinícius de Lacerda, "needed an escort to prove that chloroquine does not work", in addition, the researchers associated with his studies "began to face a wave of lynching on social media, with threats and personal attacks" [http://informe.ensp.fiocruz.br/noticias/51215]. With the growing number of conspiracy theories surrounding the COVID-19 Pandemic and the institutional support of the |

| | Federal Government that led to the population's adherence to "early treatment", science and the scientific method were notably attacked on a daily basis - thus implicating form, cyclically on public policies themselves, which are distanced from evidence and legitimized by rhetoric that mixes Science with "point of view". |
|---|---|

Source: Own elaboration (2024).

## 6. Authors biography

**Ergon Cugler de Moraes Silva** has a Master's degree in Public Administration and Government (FGV), Postgraduate MBA in Data Science & Analytics (USP) and Bachelor's degree in Public Policy Management (USP). He is associated with the Bureaucracy Studies Center (NEB FGV), collaborates with the Interdisciplinary Observatory of Public Policies (OIPP USP), with the Study Group on Technology and Innovations in Public Management (GETIP USP) with the Monitor of Political Debate in the Digital Environment (Monitor USP) and with the Working Group on Strategy, Data and Sovereignty of the Study and Research Group on International Security of the Institute of International Relations of the University of Brasília (GEPSI UnB). He is also a researcher at the Brazilian Institute of Information in Science and Technology (IBICT), where he works for the Federal Government on strategies against disinformation. São Paulo, São Paulo, Brazil. Web site: https://ergoncugler.com/.

**José Carlos Vaz** is a Professor at the University of São Paulo - School of Arts, Sciences, and Humanities, in the undergraduate and graduate courses in Public Policy Management. Vice-president of the Administrative Council of the Institute Pólis. Coordinator of GETIP - Study Group on Technology and Innovation in Public Management. Bachelor's degree in Administration from the University of São Paulo (1986), Master's degree in Public Administration from Fundação Getulio Vargas EAESP (1995), and a Ph.D. in Business Administration - Information Systems from Fundação Getúlio Vargas EAESP (2003). Has experience in the field of Public Administration, focusing mainly on the following themes: social and political aspects of information technology use (digital participation, open government data, social control of governments, e-government, technology public policies), public management (state and government capabilities, innovations in public management, logistics, strategic planning), and urban and municipal issues (urban processes and dynamics, urban mobility, municipal management, local development).

-

**Note:** Data survey initially conducted through the Undergraduate Thesis titled "How disinformation impacts public policies?: Elements for constructing an analytical model to observe potential impacts of disinformation and fake news on public policies" (Jan. 2022), from the Public Policy Management course at the School of Arts, Sciences, and Humanities (EACH) of the University of São Paulo (USP), defended by Ergon Cugler de Moraes Silva, under the supervision of Professor Dr. José Carlos Vaz, with Professor Dr. Cristiane Kerches da Silva Leite and Professor Dr. Márcio Moretto Ribeiro on the defense committee. Additionally, the institutional booklet "How disinformation impacts public policies" (Sept. 2022) was developed with the support of the Study Group on Technology and Innovations in Public Management (GETIP) and the Interdisciplinary Observatory of Public Policies "Professor Dr. José Renato de Campos Araújo" (OIPP).

# Como a desinformação e as *fake news* impactam as políticas públicas?: Uma revisão da literatura internacional [Português]

*Ergon Cugler de Moraes Silva*

Fundação Getulio Vargas (FGV)
Universidade de São Paulo (USP)
São Paulo, São Paulo, Brasil

contato@ergoncugler.com
www.ergoncugler.com

*José Carlos Vaz*

Universidade de São Paulo (USP)
Fundação Getulio Vargas (FGV)
São Paulo, São Paulo, Brasil

vaz@usp.br
www.getip.net.br

**Resumo**

Este estudo investiga o impacto da desinformação nas políticas públicas. Utilizando 28 conjuntos de palavras-chave em oito bases de dados, foi realizada uma revisão sistemática seguindo o modelo "Prisma 2020" (Page et al., 2021). Após aplicar filtros e critérios de inclusão e exclusão sobre 4.128 artigos e materiais inicialmente encontrados, 46 publicações foram analisadas, resultando em 23 categorias de impacto da desinformação. Essas categorias foram organizadas em dois eixos principais: "Estado e Sociedade" e "Atores e Dinâmicas", abrangendo impactos sobre atores do Estado, atores da sociedade, dinâmicas do Estado e dinâmicas da sociedade. Os resultados indicam que a desinformação afeta decisões públicas, adesão a políticas, prestígio das instituições, percepção da realidade, consumo, saúde pública e outros aspectos. Além disso, este estudo sugere que a desinformação deve ser tratada como um problema público e incorporada à agenda de pesquisa em políticas públicas, contribuindo para o desenvolvimento de estratégias para mitigar seus efeitos nas ações governamentais.

## 1. Introdução

Este estudo tem como objetivo investigar como a desinformação afeta as políticas públicas. Para isso, foram selecionados 28 conjuntos de palavras-chave, aplicados em oito bases de dados diferentes, resultando em uma revisão sistemática baseada no modelo "Prisma 2020" (Page et al., 2021). Após duas rodadas de aplicação de filtros e critérios rigorosos de inclusão e exclusão, foram identificadas 46 publicações relevantes. Essas publicações foram agrupadas por temas e analisadas, culminando na definição de 23 categorias analíticas para observar os impactos da desinformação nas políticas públicas.

Para aprofundar a análise, essas 23 categorias foram organizadas em dois eixos principais: "Estado" e "Sociedade", além de "Atores" e "Dinâmicas". Esse enquadramento permitiu uma observação detalhada dos impactos da desinformação em quatro áreas distintas: **a) impactos sobre atores do Estado; b) impactos sobre atores da sociedade; c) impactos sobre dinâmicas do Estado; e d) impactos sobre dinâmicas da sociedade.**

**Impactos sobre atores do Estado:** Foi observado que a desinformação pode impactar significativamente os atores do Estado, incluindo agentes do judiciário, reguladores e burocratas em diferentes níveis. Por exemplo, agentes do judiciário podem tomar decisões com base em informações falsas, comprometendo o acesso à Justiça e a credibilidade das Instituições. Um caso notável envolve decisões judiciais influenciadas por teorias da conspiração ou notícias falsas, que podem resultar em sentenças equivocadas e perda de confiança pública no sistema judicial. Além disso, órgãos reguladores, ao basear suas políticas em dados incorretos, podem implementar regulamentações inadequadas que afetam setores críticos da economia e a segurança pública.

**Impactos sobre atores da sociedade:** Também foi observado que a desinformação afeta atores importantes da sociedade, como a mídia, organizações da sociedade civil e o setor privado. A mídia, por exemplo, pode incorrer em informações falsas, influenciando a opinião pública e direcionando o debate público para temas enganosos. Uma disputa narrativa se evidenciou durante a pandemia de COVID-19, por exemplo, onde a disseminação de informações falsas sobre tratamentos e vacinas afetou a adesão da população às medidas de saúde pública. Organizações da sociedade civil podem ser enganadas por dados falsos, comprometendo suas campanhas e esforços de *advocacy*. Empresas privadas, ao tomarem decisões baseadas em desinformação, podem enfrentar prejuízos financeiros significativos e contribuir para a desestabilização econômica.

**Impactos sobre dinâmicas do Estado:** Nas dinâmicas do Estado, a desinformação pode afetar a comunicação institucional e o planejamento de políticas públicas. A comunicação institucional é crucial para garantir que informações corretas cheguem à população. Quando a desinformação prevalece, observa-se que o Estado precisa investir mais em campanhas de esclarecimento, como as observadas durante campanhas de vacinação, onde foi necessário combater mitos e desinformações. Isso representa um desvio de recursos que poderiam ser utilizados em outras áreas essenciais. Além disso, a desinformação pode comprometer o planejamento público, levando a uma má alocação de recursos e à implementação ineficaz de políticas, como evidenciado em crises de saúde pública onde a resposta estatal foi prejudicada por informações incorretas.

**Impactos sobre dinâmicas da sociedade:** A desinformação pode alterar também significativamente a percepção da realidade e a adesão da população a políticas públicas. Por exemplo, informações falsas sobre a segurança das vacinas podem reduzir a taxa de vacinação, resultando em surtos de doenças preveníveis. Isso foi observado com o ressurgimento de doenças como o sarampo em regiões onde movimentos antivacina ganharam força. Além disso, a desinformação pode polarizar a opinião pública, reforçando crenças incorretas e dificultando o consenso em questões políticas e sociais importantes. Essa polarização pode levar a conflitos sociais e dificultar a implementação de políticas públicas baseadas em evidências, prejudicando o bem-estar coletivo e a coesão social.

Desta forma, este estudo sugere que a desinformação deve ser reconhecida como um problema público significativo e incorporada à agenda de pesquisa em políticas públicas. Entender e mitigar os impactos da desinformação pode contribuir para o desenvolvimento de estratégias mais eficazes na formulação e implementação de políticas públicas, fortalecendo assim a governança e a confiança nas instituições públicas.

## 2. Materiais e métodos

A partir da "revisão sistemática integrativa" proposta por Botelho et. al (2011, p. 127), foi realizada a busca em títulos e resumos de publicações, sem restrição temporal, utilizando 28 buscas com operador booleano 'AND'. A Figura 01 detalha o fluxo.

**Figura 01.** Modelo "Prisma 2020 Fluxogram" para revisão sistemática de literatura:

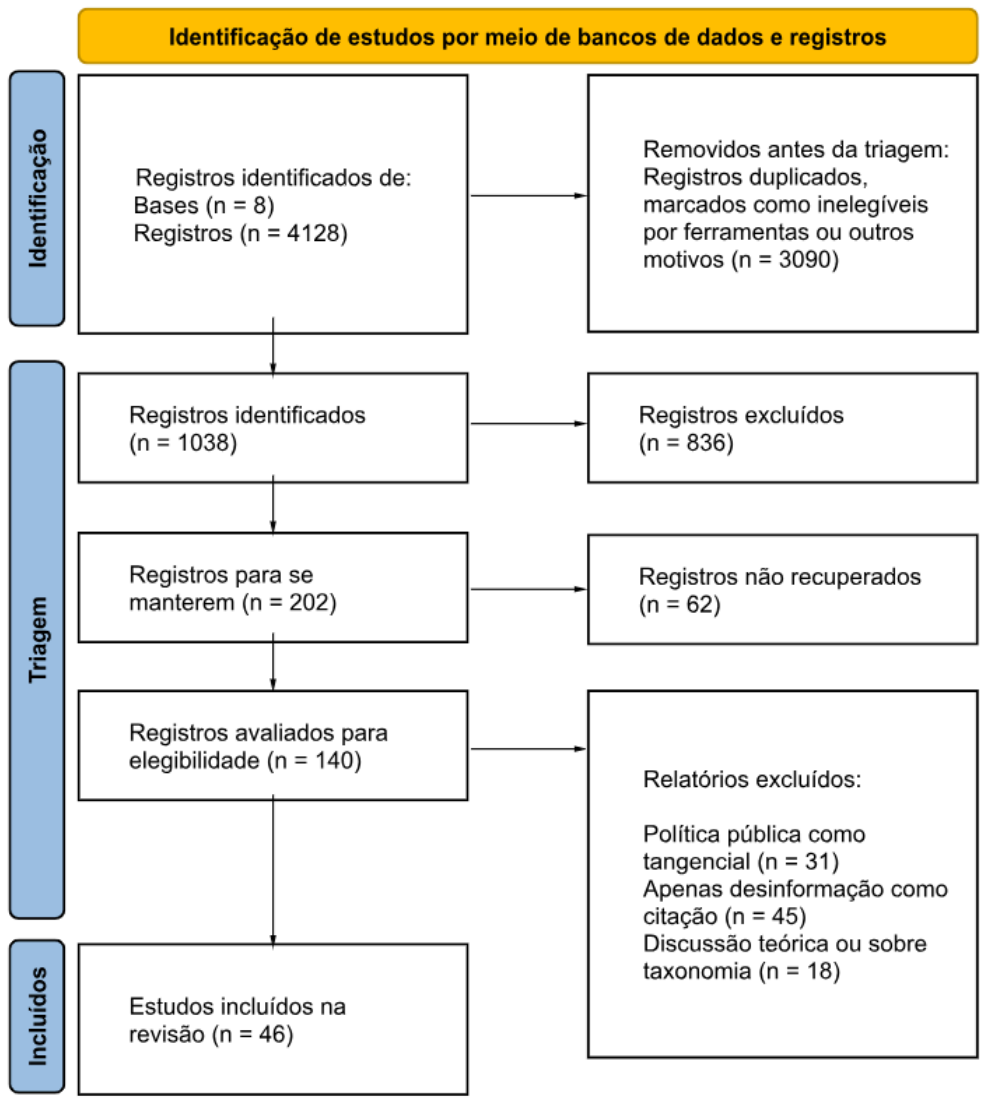

Fonte: Elaboração própria (2024).

Foram elencadas as palavras-chave, sendo em português: "fake news" + "literatura"; "desinformação" + "literatura"; "agnotologia" + "literatura"; "fake news" + "efeito"; "desinformação" + "efeito"; "agnotologia" + "efeito"; "fake news" + "impacto"; "desinformação" + "impacto"; "agnotologia" + "impacto"; "fake news" + "políticas públicas"; "desinformação" + "políticas públicas"; "agnotologia" + "políticas públicas". Em inglês: "fake news" + "literature review"; "misinformation" + "literature review"; "disinformation" + "literature review"; "agnotology" + "literature review"; "fake news" + "effect"; "misinformation" + "effect"; "disinformation" + "effect"; "agnotology" + "effect"; "fake news" + "impact"; "misinformation" + "impact"; "disinformation" + "impact"; "agnotology" + "impact"; "fake news" + "public policy"; "misinformation" + "public policy"; "disinformation" + "public policy"; "agnotology" + "public policy".

Utilizou-se ainda da proposta do método "Prisma 2020" (Page et al., 2021), sendo um conjunto de passos baseados em evidências para conduzir relatórios em revisões sistemáticas. Tais 28 conjuntos de palavras-chave foram buscadas em 8 bases de dados: 1.) Google Scholar; 2.) ProQuest; 3.) Scielo (Scientific Electronic Library Online); 4.) Portal de Periódicos Capes; 5.) Catálogo de Teses & Dissertações da Capes; 6.) Portal da Biblioteca Digital de Teses e Dissertações da Universidade de São Paulo (USP); 7.) Oasis BR Ibict; e 8.) Biblioteca Digital Brasileira de Teses e Dissertações (BDTD). Um levantamento inicial trouxe um total de 4.128 resultados com palavras-chaves no título ou resumo e, de forma sistematizada, foram selecionados 46 estudos para serem revisados.

## 3. Explorando os resultados

**Quadro 01.** Categorias de análise:

| Categoria | Descrição | Referencial utilizado |
|---|---|---|
| Impactos sobre **agentes do Judiciário** | Impactos sobre atores do Estado alocados no Judiciário e suas tomadas de decisões e influência em relação às políticas públicas e ações governamentais (ex: decisões públicas baseadas em desinformação; ou omissões diante de desinformação; ou ainda posições sustentadas por sistemas de crenças no contraponto do conhecimento científico). | **1.** Oriras et al., 2018; **2.** Castinholi, 2019; **3.** Araujo et al., 2020. |
| Impactos sobre **agentes reguladores e fiscalizadores** | Impactos sobre atores do Estado alocados em órgãos reguladores e fiscalizadores (ex: estatais, agências e outros) e suas tomadas de decisões e influência em relação às políticas públicas e ações governamentais (ex: decisões públicas baseadas em desinformação; ou omissões diante de desinformação; ou ainda posições sustentadas por sistemas de crenças no contraponto do conhecimento científico). | **1.** Oriras et al., 2018; **2.** Lewandowsky, 2021; **3.** Soares, 2020; **4.** Vignoli et al., 2021; **5.** Castinholi, 2019. |
| Impactos sobre **burocratas de médio escalão** | Impactos sobre atores do Estado alocados na burocracia de médio escalão (ex: gerentes de departamento e outros) e suas tomadas de decisões e influência em relação às políticas públicas e ações governamentais (ex: decisões | **1.** Rodríguez-Fernández, 2019; **2.** Soares, 2020; **3.** Keenan et al., 2018; **4.** Vignoli et al., 2021. |

| Categoria | Descrição | Referencial utilizado |
|---|---|---|
| | públicas baseadas em desinformação; ou omissões diante de desinformação; ou ainda posições sustentadas por sistemas de crenças no contraponto do conhecimento científico). | |
| Impactos sobre **burocratas de nível de rua** | Impactos sobre atores do Estado alocados na burocracia de nível de rua (ex: policiais, professores, profissionais da saúde e outros) e suas tomadas de decisões e influência em relação às políticas públicas e ações governamentais (ex: decisões públicas baseadas em desinformação; ou omissões diante de desinformação; ou ainda posições sustentadas por sistemas de crenças no contraponto do conhecimento científico). | **1.** Keenan; Dillenburger, 2018. |
| Impactos sobre **burocratas policymakers** | Impactos sobre atores do Estado alocados na burocracia de alto escalão (ex: secretários, ministros, diretores de estatais, formuladores com incidência no design de políticas públicas) e suas tomadas de decisões e influência em relação às políticas públicas e ações governamentais (ex: decisões públicas baseadas em desinformação; ou omissões diante de desinformação; ou ainda posições sustentadas por sistemas de crenças no contraponto do conhecimento científico). | **1.** Rodríguez-Fernández, 2019; **2.** Soares, 2020; **3.** Vignoli et al., 2021; **4.** Castinholi, 2019; **5.** Araujo et al., 2020. |
| Impactos sobre **conselhos institucionais e comissões** | Impactos sobre atores do Estado e atores da sociedade que compõem comissões eletivas e conselhos institucionais e suas tomadas de decisões e influência em relação às políticas públicas e ações governamentais (ex: decisões públicas baseadas em desinformação; ou omissões diante de desinformação; ou ainda posições sustentadas por sistemas de crenças no contraponto do conhecimento científico). | **1.** Rodríguez-Fernández, 2019; **2.** Soares, 2020; **3.** Vignoli et al., 2021, **4.** Araujo et al., 2020. |
| Impactos sobre **parlamentares** | Impactos sobre atores do Estado com função eletiva e representativa no Legislativo e suas tomadas de decisões e influência em relação às políticas públicas e ações governamentais (ex: decisões públicas baseadas em desinformação; ou omissões diante de desinformação; ou ainda posições sustentadas por sistemas de crenças no contraponto do conhecimento científico). | **1.** Lewandowsky, 2021; **2.** Vignoli et al., 2021; **3.** Araujo et al., 2020. |
| Impactos sobre a **mídia e imprensa** | Impactos sobre a mídia e a imprensa e influência em relação às políticas públicas e | **1.** Pate et al., 2019; **2.** Lisboa et al., 2020; **3.** Mello, 2020. |

| Categoria | Descrição | Referencial utilizado |
|---|---|---|
| | ações governamentais (ex: posições de setores diante de pautas e políticas públicas; mobilização de pauta jornalística em torno de uma agenda com busca de aprovação). | |
| Impactos sobre **organizações da sociedade civil** | Impactos sobre atores da sociedade em Organizações da sociedade civil e influência em relação às políticas públicas e ações governamentais (ex: decisões públicas baseadas em desinformação; ou omissões diante de desinformação; ou ainda posições sustentadas por sistemas de crenças no contraponto do conhecimento científico). | **1.** Rodríguez-Fernández, 2019; **2.** Vignoli et al., 2021; **3.** Araujo et al., 2020. |
| Impactos sobre **partidos e comunidades epistêmicas** | Impactos sobre atores do Estado e da sociedade que compõem partidos políticos e/ou comunidades epistêmicas e influência em relação às políticas públicas e ações governamentais (ex: decisões públicas baseadas em desinformação; ou omissões diante de desinformação; ou ainda posições sustentadas por sistemas de crenças no contraponto do conhecimento científico). | **1.** Rodríguez-Fernández, 2019; **2.** Vignoli et al., 2021, **3.** Araujo et al., 2020. |
| Impactos sobre o **setor privado e corporações** | Impactos sobre o setor privado e influência em relação às políticas públicas e ações governamentais (ex: posições de corporações diante de pautas e políticas públicas; mobilização do setor empresarial em torno de uma agenda com busca de aprovação). | **1.** Rodríguez-Fernández, 2019. |
| Impactos sobre a **comunicação institucional e TICs** | Impactos sobre a estrutura do Estado em relação à comunicação institucional, relações públicas e/ou demandas por capacidades estatais de práticas e Tecnologias de Informação e Comunicação para contrapor desinformação (ex: propagandas anti-*fake news*; ou tecnologias de *fact checking*). | **1.** Carneiro, 2019; **2.** Silva, 2017; **3.** Vaccari et al., 2020; **4.** Whyte, 2020; **5.** Matasick et al., 2020; **6.** Vignoli et al., 2021. |
| Impactos sobre o **orçamento público e despesas** | Impactos sobre a estrutura do Estado em relação ao orçamento público e despesas (ex: despesas de atividades de enfrentamento à desinformação; ou mesmo perda de bens que levem a custos ao Estado). | **1.** Vaccari et al., 2020; **2.** Whyte, 2020; **3.** Matasick et al., 2020. |
| Impactos sobre o **planejamento público e recursos** | Impactos sobre a estrutura do Estado em relação ao planejamento público (ex: metas, previsibilidade, gestão de crises) e alocação de recursos e insumos (ex: recursos humanos, | **1.** Matasick et al., 2020. |

| Categoria | Descrição | Referencial utilizado |
|---|---|---|
| | materiais, ou demandas estatais). | |
| Impactos sobre o **prestígio das Instituições e do Sistema** | Impactos sobre a estrutura do Estado em relação ao prestígio das Instituições e da democracia (ex: discursos de ódio; fragilidade das instituições; desmobilização estatal; ou ataques à democracia). | **1.** Rodríguez-Fernández, 2019; **2.** Milla et al., 2020; **3.** Pate et al., 2019; **4.** Silva, 2019; **5.** Patihis et al., 2018; **6.** Benedict et al., 2019; **7.** Greenspan et al., 2020; **8.** Prandi et al., 2020; **9.** Ferreira et al., 2021; **10.** Than et al., 2020; **11.** Ghenai et al., 2017; **12.** McNamara, 2019; **13.** Miller, 2019; **14.** Newton, 2019; **15.** Huang et al., 2019; **16.** Souza Junior et al., 2020; **17.** Souza, 2020; **18.** Patel., 2020; **19.** Araujo et al., 2020; **20.** Da Empoli, 2019; **21.** Norris et al., 2019; **22.** Benkler et al., 2018; **23.** Alves, 2020. |
| Impactos sobre as **relações domésticas e/ou relações exteriores** | Impactos sobre a estrutura do Estado no relacionamento entre entes (ex: entre esferas; entre os Poderes; entre entes exteriores; entre povos; ou organizações multilaterais globais). | **1.** Cadwalladr et al., 2018; **2.** Valente, 2019; **3.** Pate et al., 2019; **4.** Mejias et al., 2017; **5.** La Cour, 2020; **6.** Landon-Murray et al., 2019. |
| Impactos sobre a **adesão à políticas públicas** | Impactos sobre a estrutura da sociedade no que diz respeito à adesão à políticas públicas por parte dos cidadãos (ex: adesão às campanhas de vacinação; ou às orientações estatais; alinhamento de pautas nacionais). | **1.** Keenan et al., 2018; 2. Pate et al., 2019; **3.** Prandi et al., 2020; **4.** Silva et al., 2017; **5.** Guimarães, 2017, apud Silva et al., 2017; **6.** Vignoli et al., 2021; **7.** Than et al., 2020; **8.** Ghenai et al., 2017; **9.** McNamara, 2019; **10.** Miller, 2019; **11.** Newton, 2019; **12.** Huang et al., 2019; **13.** Souza Junior et al., 2020; **14.** Souza, 2020, apud Almeida et al., 2020. |
| Impactos sobre as **alterações na cadeia de produção** | Impactos sobre a estrutura da sociedade no que diz respeito às cadeias de produção, inclusive na relação de tais cadeias com fluxos logísticos de produção do Estado (ex: paralisações/greves de cadeias de produção). | **1.** Almeida et al., 2020; **2.** Zhang, 2020, apud Almeida et al., 2020. |
| Impactos sobre as **alterações na percepção da realidade** | Impactos sobre a estrutura da sociedade no que diz respeito às alterações, inclusive psicológicas, na percepção da realidade (ex: narrativas em disputa da verdade; conflitos diante de orientações ambíguas; ou ainda impactos psicológicos e da memória em relação à percepção da realidade). | **1.** Silva, 2019; **2.** Patihis et al., 2018; **3.** Benedict et al., 2019; **4.** Greenspan et al., 2020; **5.** Pate et al., 2019; **6.** Almeida et al., 2020; **7.** Peeri et al., 2020; **8.** Santos et al., 2020; **9.** Matos et al., 2020; **10.** Neto et al., 2020; **11.** Sutherland et al., 2001; **12.** Wylie et al., 2014, apud Greenspan et al., 2020; **13.** Vignoli et al., 2021; **14.** Patel., 2020; **15.** Araujo et al., 2020; **16.** Da Empoli, 2019; **17.** Norris et al., |

| Categoria | Descrição | Referencial utilizado |
|---|---|---|
| | | 2019; **18.** Benkler et al., 2018; **19.** Alves, 2020. |
| Impactos sobre as **alterações no consumo** | Impactos sobre a estrutura da sociedade no que diz respeito às alterações em padrões sociais de consumo de bens e serviços (ex: consumo de medicamentos sem prescrição de especialistas; ou até assimetrias no consumo de insumos e alimentos, tal como casos de rumores sobre desabastecimentos). | **1.** Almeida et al., 2020; **2.** Zaracostas, 2019, apud Almeida et al., 2020. |
| Impactos sobre **externalidades, integridade e ambiente** | Impactos sobre indivíduos e/ou grupos sociais em relação à sua saúde, integridade e/ou segurança (ex: fatalidades ou enfermidades que resultaram de atos de desinformações; externalidades ao meio ambiente, sociais e/ou econômicas; ou ainda externalidade setorial relacionada à política pública observada.). | **1.** Lisboa et al., 2020; **2.** Libório et al., 2020, apud Lisboa et al., 2020; **3.** Islam et al., 2020; **4.** Valente, 2019; **5.** Pate et al., 2019; **6.** Mejias et al., 2017; **7.** Almeida et al., 2020; **8.** Peeri et al., 2020; **9.** Santos et al., 2020; **10.** Matos et al., 2020; **11.** Neto et al., 2020; **12.** Zaracostas, 2019; **13.** Than et al., 2020; **14.** Ghenai et al., 2017; **15.** McNamara, 2019; **16.** Miller, 2019; **17.** Newton, 2019; **18.** Huang et al., 2019; **19.** Souza Junior et al., 2020; **20.** Souza, 2020, apud Almeida et al., 2020. |
| Impactos sobre a **opinião pública e sistema de crenças** | Impactos sobre a estrutura da sociedade no que diz respeito à alteração e construção da opinião pública e do imaginário social em meio aos sistemas de crenças em disputa (ex: discriminação em relação à povos e/ou grupos sociais específicos; ou narrativas taxativas sobre algum programa e política pública; ou ainda a própria percepção e opinião pública sobre determinada política pública). | **1.** Oriras et al., 2018; **2.** Milla et al., 2020; **3.** Pate et al., 2019; **4.** Sutherland et al., 2001; **5.** Wylie et al., 2014, apud Greenspan et al., 2020; **6.** Vignoli et al., 2021; **7.** Ferreira et al., 2021; **8.** Castinholi, 2019; **9.** Patel, 2020; **10.** Araujo et al., 2020; **11.** Da Empoli, 2019; **12.** Norris et al., 2019; **13.** Benkler et al., 2018; **14.** Alves, 2020. |
| Impactos sobre o **prestígio do conhecimento científico** | Impactos sobre a estrutura da sociedade no que diz respeito ao prestígio da Ciência, inclusive da comunidade científica em suas diversas áreas (ex: teorias da conspiração; narrativas ambíguas desvalidando a Ciência; desmonte científico legitimado por retórica). | **1.** Almeida et al., 2020; **2.** Peeri et al., 2020; **3.** Santos et al., 2020; **4.** Matos et al., 2020; **5.** Neto et al., 2020; **6.** Sutherland et al., 2001; **7.** Wylie et al., 2014, apud Greenspan et al., 2020; **8.** Silva et al., 2017; **9,** Vignoli et al., 2021; **10.** Patel., 2020; **11.** Araujo et al., 2020; **12.** Da Empoli, 2019; **13.** Norris et al., 2019; **14.** Benkler et al., 2018; **15.** Alves, 2020. |

Fonte: Elaboração própria. *A partir de autores revisados em sua publicação (2024).

Além disso, a Figura 02 a seguir ilustra os 23 potenciais impactos da desinformação:

**Figura 02.** Matriz de impactos da desinformação:

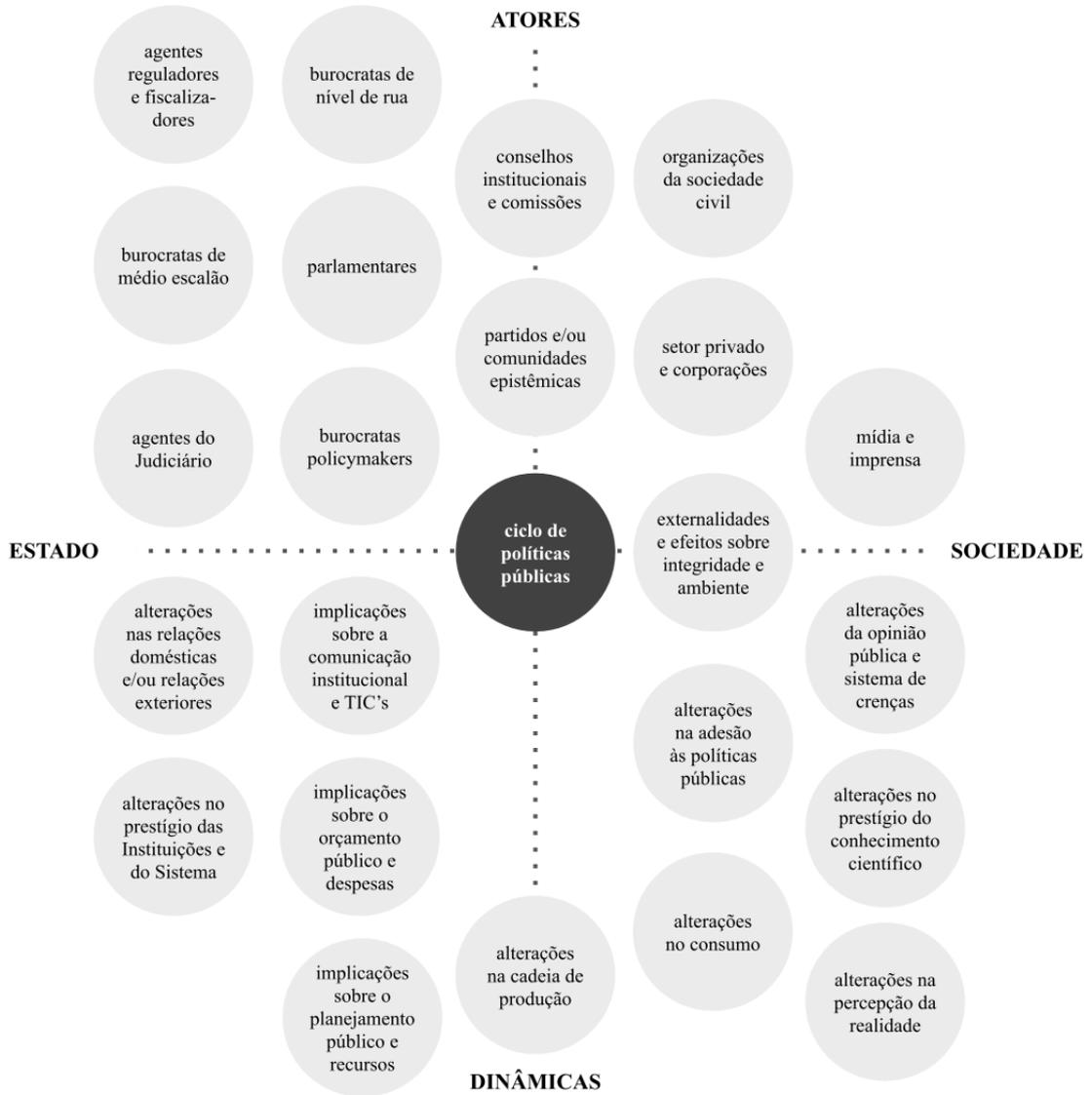

Fonte: Elaboração própria (2023).

## 4. Experimentação do modelo analítico

**Quadro 02.** Teste com caso do tratamento precoce irregular (*off label*) na Pandemia da COVID-19:

| **Impactos sobre Atores do Estado** | |
| --- | --- |
| **Categoria** | **Impactos observados** |
| Impactos sobre **agentes do Judiciário** | As discussões sobre Hidroxi/Cloroquina geraram tumulto entre juristas. A Defensoria Pública da União (DPU) chegou a mover ação contra o Conselho Federal de Medicina (CFM) diante da omissão deste sobre o suposto tratamento precoce. Setores críticos chegaram a atacar publicamente a posição da DPU [https://revistaoeste.com/politica/mais-uma-aberracao-do-sistema-judiciario/], enquanto o Conselho Nacional de Justiça (CNJ) precisou emitir parecer para orientar Juízes [https://www.conjur.com.br/2020-mar-23/cnj-divulga-parecer-orientar-juizes-hidroxicloroquina] diante das ambiguidades sobre o tema. Em suma, a desinformação sobre o tratamento precoce, além de custar esforços e atenção do Poder Judiciário, levou a decisões locais que permitiram a legitimação do tratamento precoce com previsão na autonomia médica [https://www.conjur.com.br/2020-mai-30/responsabilizacao-juridica-uso-cloroquina-nao-consenso]. Consequente da desinformação e de decisões desalinhadas do então entendimento da comunidade científica, tem-se a política de enfrentamento à Pandemia da COVID-19 dispersada, com varas dependentes de suas respectivas decisões judiciais em relação ao tema, até que o STF viesse a posicionar-se oficialmente sobre [https://www.cnnbrasil.com.br/politica/stf-determina-que-governo-bolsonaro-apresente-protocolo-para-tratamento-da-covid/], buscando então pacificar a desinformação que estava criando ambiguidades entre colegiados e suas decisões. Destaca-se um impacto sobre a implementação da política sanitária, dispersando orientações e normativas. |
| Impactos sobre **agentes reguladores e fiscalizadores** | Não foram encontrados impactos da desinformação do "tratamento precoce" sobre agentes e órgãos reguladores. Pelo contrário, logo no início da Pandemia (abril de 2020), a Agência Nacional de Vigilância Sanitária (ANVISA) havia apontado ao Governo Bolsonaro que a Hidroxi/Cloroquina não deveria ser usada contra a COVID-19 [https://www.brasildefato.com.br/2020/08/24/em-abril-anvisa-alertou-bolsonaro-que-cloroquina-nao-deveria-ser-usada-contra-covid]. |
| Impactos sobre | Vale luz sobre um caso envolvendo os burocratas de médio |
| **burocratas de médio escalão** | escalão do Ministério da Saúde (MS), no qual tais burocratas, independente de quais sejam suas motivações (as quais não cabem a esta discussão), mas estando envolvidos na pauta do "tratamento precoce", passaram a obstruir as investigações da Comissão Parlamentar de Inquérito (CPI) da COVID-19 no Senado. Como respostas, a CPI votou pelo afastamento de tais servidores [https://oglobo.globo.com/politica/cpi-da-covid-votara-afastamento-de-servidores-que-estao-obstruindo-investigacoes-ou-poderiam-obstruir-1-25131794]. Vale citar que o MS conta com diversos cargos estratégicos loteados por atores envolvidos na pauta do "tratamento precoce". Em posições de chefia, Hélio Angotti e Mayra Pinheiro militam ativamente mesmo após quase dois anos de Pandemia, em defesa do "Kit Covid" [https://www.correiobraziliense.com.br/politica/2021/02/4904951-mais-de-20-militares----capita-cloroquina----e-medico-olavista-ocupam-cargos-no-ministerio-da-saude.html] e assim, consequentemente, carregam uma pauta sem quaisquer evidências científicas para a rotina do MS, incidindo diretamente sobre políticas públicas. Dado o contexto de circulação dos atores de médio escalão, observa-se como estes podem impactar em todas etapas do ciclo de políticas públicas quando observada a disputa do "tratamento precoce", pois desde a formulação do "TrateCov", até sua implementação e tentativa de distorção da avaliação em meio à investigação, tais burocratas exerceram influência. |
| Impactos sobre **burocratas de nível de rua** | Mesmo um ano após o início da Pandemia da COVID-19, levantamento realizado pela Associação Médica Brasileira (AMB) mostrou que dentre os médicos que estão atuando na linha de frente, 37,4% acreditam que a cloroquina é eficaz para prevenir a Covid-19, enquanto 41,4% veem o mesmo potencial na ivermectina [https://olhardigital.com.br/2021/02/03/videos/covid-19-1-3-dos-medicos-acredita-em-tratamento-precoce/] mesmo sem nenhuuma evidência científica para tal. Em pesquisa coordenada pela professora Gabriela Lotta (FGV), na qual 1.520 Agentes Comunitários de Saúde (ACS's) são entrevistados, apesar de que "mais de 90% dos entrevistados desaprovam, total ou parcialmente, que o presidente Jair Bolsonaro prescreva medicamentos para a população", observa-se que um terço (33,8%) dos ACS's entrevistados não acham que apenas drogas com comprovação de eficácia devem ser utilizadas, abrindo brecha para discricionariedades em meio às discussões sobre "tratamento precoce" [https://www.revistaquestaodeciencia.com.br/artigo/2020/11/11/pesquisa-da-fgv-mostra-falta-de-treinamento-de-testes-e-medo-entre-profissionais-de-saude-no-sus]. Cabe destaque não tão somente aos impactos na implementação da política |
| | pública, mas também nas vulnerabilidades que as desigualdades trazem diante das assincronias de informações e as distorções impulsionadas. Comunidades dependentes da capilaridade do SUS, por exemplo, contam com a atuação de ACS's para praticar sua relação com o Estado. |
| Impactos sobre **burocratas policymakers** | Entre os conflitos e ambiguidades mais evidentes em relação ao tratamento precoce, os próprios burocratas de alto escalão do Governo Federal destacam-se na disputa interna e com demais esferas e Poderes do país. Internamente, foram dois Ministros da Saúde demitidos após não terem se posicionado publicamente em favor do suposto "tratamento precoce" com Hidroxi/Cloroquina [https://www1.folha.uol.com.br/cotidiano/2020/12/tres-ministros-da-saude-e-uma-pandemia-o-ano-em-que-ficamos-doentes.shtml], apesar da inexistência de qualquer evidência ou mesmo organização científica internacional apoiando tal tese. Diante de todas documentações que evidenciam a omissão e descoordenação Federal [https://www.camara.leg.br/noticias/766207-instituicoes-sociais-denunciam-omissao-do-governo-no-combate-a-pandemia], não restam dúvidas sobre como a desinformação pautou atores políticos do alto escalão, incidindo diretamente em propostas como a do Ministério da Saúde (MS) sobre o "Kit Covid", baseado na falsa tese de "tratamento precoce" e "imunidade de rebanho". Foram apenas meses após que o próprio MS assumiu a ineficácia do próprio "Kit Covid" [https://pebmed.com.br/ministerio-da-saude-confirma-ineficacia-do-kit-covid-no-tratamento-contra-covid-19/]. Neste sentido, observa-se como a desinformação circula em meio aos formuladores, passando desta etapa do ciclo de políticas públicas para a práxis da implementação. |
| Impactos sobre **conselhos institucionais e comissões** | Considerando Conselhos Institucionais e Comissões com incidência em políticas públicas relacionadas ao tema da Saúde e da gestão da Pandemia da COVID-19, não foram encontrados impactos da desinformação sobre estas. Pelo contrário, o Conselho Nacional de Saúde (CNS) chegou a protocolar documentos em denúncia junto à Comissão Parlamentar de Inquérito (CPI) da COVID-19 no Senado junto a demais organizações e entes [https://www.camara.leg.br/noticias/766207-instituicoes-sociais-denunciam-omissao-do-governo-no-combate-a-pandemia]. Além disso, foi o próprio CNS que "pediu [para] que o Ministério da Saúde revogasse a nota técnica que orientava o uso da cloroquina e se abstivesse de incentivar o uso de medicamentos sem eficácia e segurança comprovada" [https://www.nexojornal.com.br/expresso/2021/05/24/Qual-a-cronologia-cient%C3%ADfica-da-cloroquina-na-pandemia]. Em meio à disputa de formulação das diretrizes das políticas sanitárias, tal etapa do ciclo de políticas públicas |
| Impactos sobre **parlamentares** | também se vulnerabiliza.<br><br>Parlamentares Federais da base de apoio do Governo, em consonância com as teses de "imunidade de rebanho" e "tratamento precoce", chegaram a apresentar Projeto de Lei para autorizar a distribuição de Hidroxi/Cloroquina [https://www25.senado.leg.br/web/atividade/materias/-/materia/141533], as discussões da Comissão Parlamentar de Inquérito (CPI) da COVID-19 no Senado não deixaram dúvidas sobre como a Hidroxi/Cloroquina foi adotada como plataforma dos parlamentares apoiadores do Governo Bolsonaro, levando inclusive o presidente da CPI a apresentar Projeto de Lei na antítese: proibindo e criminalizando o incentivo ao uso de remédios sem comprovação científica [https://oglobo.globo.com/brasil/presidente-da-cpi-apresenta-projeto-que-criminaliza-incentivo-ao-uso-de-remedios-sem-comprovacao-cientifica-25028819]. Em suma, além dos conflitos e ambiguidades gerados pela desinformação acerca do "tratamento precoce" - este sem qualquer evidência científica até o presente momento em relação à COVID-19 -, e de existirem parlamentares baseando-se em tais teses para desenhar políticas públicas, observa-se a vulnerabilidade do Parlamento à tais disputas retóricas nesta pauta. |

**Impactos sobre Atores da Sociedade**

| Categoria | Impactos observados |
|---|---|
| Impactos sobre a **mídia e imprensa** | Com exceção de mídias alternativas ideologicamente posicionadas e com práticas de disseminação de desinformação, foi raro observar na mídia tradicional posturas institucionais e públicas de defesa do então tratamento precoce. Ainda assim, o professor Carlos Eduardo Lins da Silva (USP) ressalta que a mídia brasileira tem cometido o mesmo erro da mídia estadunidense e dando "muito espaço e destaque para comentar sobre os negacionistas, terraplanistas ou os favoráveis ao uso da cloroquina. É preciso pesar bem quais são os defensores de uma ou de outra teoria. 'Não dar o mesmo espaço na mídia para os que aprovam o uso do medicamento em detrimento das entidades científicas sérias'" [https://jornal.usp.br/radio-usp/colunista-analisa-imprensa-americana-e-brasileira-na-abordagem-da-cloroquina/]. Não se tratando da adoção de uma postura negacionista institucional, o impacto da desinformação está nesta ser tratada como "opinião" e "ponto de vista", na qual "deve ser escutada". Desta forma, o espaço cedido pela mídia tradicional impulsiona indiretamente retóricas de disseminação de desinformação. |
| Impactos sobre **organizações da sociedade civil** | Por outro lado, diversas organizações com relevância na área se posicionaram contrariamente à tese do "tratamento precoce", sendo a Associação Médica Brasileira (AMB), a Associação de Medicina Intensiva Brasileira (AMIB) e as sociedades de Infectologia, Imunologia, Pneumologia e Tisiologia [https://www.nexojornal.com.br/expresso/2021/05/24/Qual-a-cronologia-cient%C3%ADfica-da-cloroquina-na-pandemia]. Além disso, organizações foram articulando-se diante da pandemia, como é o caso da então fundada "Associação de Vítimas e Familiares de Vítimas da Covid-19 (Avico Brasil)", a qual propôs uma representação criminal contra o Governo Bolsonaro [https://www.camara.leg.br/noticias/766207-instituicoes-sociais-denunciam-omissao-do-governo-no-combate-a-pandemia]. |
| Impactos sobre **partidos e comunidades epistêmicas** | No que diz respeito às comunidades epistêmicas, alguns indivíduos da comunidade médica chegaram a ser influenciados por artigo publicado na conceituada The Lancet, o qual defendia o uso de Hidroxi/Cloroquina. Um único artigo em meio a centenas de outros refutando a tal eficácia do medicamento, foi suficiente para sustentar a opinião seletiva daqueles que buscavam alguma referência para legitimar suas próprias verdades. Três dias após, The Lancet suspendeu as pesquisas e refutou a eficácia publicada, retratando-se [https://www.nexojornal.com.br/expresso/2021/05/24/Qual-a-cronologia-cient%C3%ADfica-da-cloroquina-na-pandemia]. Figuras como Mayra Pinheiro e Nise Yamaguchi ganharam relevância em meio às decisões sanitárias, especialmente Pinheiro, por se tratar da Secretária da Saúde do Ministério da Saúde (MS) e militante ativa do "tratamento precoce". Além de Yamaguchi que foi cogitada ser Ministra da Saúde, e teria sugerido alterar a bula da Cloroquina para tentar legitimar seu uso contra COVID-19 [https://www.bbc.com/portuguese/brasil-57124296]. A Agência Lupa chegou a expor grupos de médicos que estariam divulgando e até patrocinando "informações falsas para defender tratamento ineficaz contra Covid-19" [https://piaui.folha.uol.com.br/lupa/2021/02/23/anuncio-medicos-pela-vida-covid-19/]. Em relação aos partidos políticos, nenhuma grande mobilização ou posição de destaque para além de divergências esperadas. |
| Impactos sobre o **setor privado e corporações** | Atores do setor Privado politicamente ativos se mobilizaram para arrecadar recursos e doações para o "Kit Covid" em apoio ao Governo Federal. Um caso emblemático é de Luciano Hang, dono da Rede Havan, o qual assumiu ter mobilizado doações para o "tratamento precoce", mesmo que este não tivesse quaisquer evidências científicas [https://www.nexojornal.com.br/expresso/2021/09/29/O-pap |

el-de-Luciano-Hang-na-difus%C3%A3o-de-rem%C3%A9dios-ineficazes]. Neste sentido, grupos de empresários buscaram incidir na condução da Pandemia da COVID-19 embarcando na desinformação da tese da Hidroxi/Cloroquina.

**Impactos sobre a Estrutura do Estado**

| Categoria | Impactos observados |
|---|---|
| Impactos sobre a **comunicação institucional e TICs** | Diante das contradições por apoiar institucionalmente o "tratamento precoce" sem quaisquer evidências científicas, o Ministério da Saúde (MS) após ser contrariado publicamente por divulgar links recomendando uso de Hidroxi/Cloroquina simplesmente apagou a publicação [https://www1.folha.uol.com.br/equilibrioesaude/2021/05/apos-reportagem-ministerio-da-saude-apaga-links-com-prescricao-de-cloroquina-para-covid.shtml]. Curiosamente, no caso observado não é uma desinformação que "bagunça" a Comunicação Institucional e faz esta ter que dar respostas à sociedade; mas o oposto, é a desinformação em posse do MS que desarticula a Comunicação Institucional ao banalizar tais canais à conteúdos sem validade científica. Neste caso, a desinformação implica sobre a Comunicação Institucional, na medida em que esta é vulgarizada de dentro para fora em prol de agendas anti-científicas. |
| Impactos sobre o **orçamento público e despesas** | Houve mobilização estatal e isenção de impostos com base na aposta da narrativa da Cloroquina. "[Foi] zerado o imposto de importação da cloroquina e da azitromicina, [com] a distribuição inicial de 3,4 milhões de unidades de cloroquina e hidroxicloroquina aos hospitais e a ampliação da produção de cloroquina pelos laboratórios químicos das Forças Armadas no Rio de Janeiro." [https://www.nexojornal.com.br/expresso/2021/05/24/Qual-a-cronologia-cient%C3%ADfica-da-cloroquina-na-pandemia]. Além do impacto evidente sobre a queda de arrecadação, cabe reforçar a mobilização das Forças Armadas para a produção em larga escala de Cloroquina. "[Foram] mais de R$ 1,5 milhão em contratos sem licitação para produzir cloroquina ou comprar insumos entre março e maio de 2020." [https://apublica.org/2021/03/o-mapa-da-cloroquina-como-governo-bolsonaro-enviou-28-milhoes-de-comprimidos-para-todo-o-brasil/], sendo inclusive alvo de investigação do Tribunal de Contas da União (TCU). Após distribuição massiva para as UF's brasileiras, 400 mil unidades seguiram em estoque e em vias para descarte [https://www.cnnbrasil.com.br/nacional/sem-demanda-nos-estados-400-mil-comprimidos-de-cloroquina-encalham-no-exercito/]. Além disso, o Governo Federal chegou a indicar R$ |
| Impactos sobre o **planejamento público e recursos** | 250 milhões para implementar o "Kit Covid" de "tratamento precoce" com distribuição para farmácias [https://www.istoedinheiro.com.br/saude-preve-gastar-r-250-milhoes-para-por-kit-covid-em-farmacias-populares/].<br><br>Com a adoção da pauta do "tratamento precoce"pelo Governo Federal, ignorou-se todo planejamento relacionado às recomendações da Organização Mundial da Saúde (OMS), tal como compra de vacinas e protocolo de vacinação [https://g1.globo.com/sp/sao-paulo/noticia/2021/01/17/apos-aprovacao-da-anvisa-governo-de-sp-aplica-1a-dose-da-coronavac-antes-do-inicio-do-plano-nacional-de-vacinacao.ghtml], ou ainda práticas de isolamento social e/ou políticas de distribuição massiva de EPI's. Neste caso, a tese do "tratamento precoce" impulsionou exatamente a omissão no planejamento para o real enfrentamento da Pandemia e, consequentemente, à descoordenação de mobilização e distribuição de recursos humanos e materiais. Não à toa, para enviar Hidroxi/Cloroquina aos estados houve celeridade [https://congressoemfoco.uol.com.br/area/governo/documento-detalha-distribuicao-de-cloroquina-a-estados/], enquanto que as vacinas foram confundidas de estados, prejudicando a cadeia logística de vacinação [https://oglobo.globo.com/saude/vacina/ministerio-da-saude-troca-remessa-de-vacinas-manda-doses-do-amazonas-para-amapa-24897507]. Um ano após o início da Pandemia, o programa de combate a malária chegou a ficar sem estoque de Cloroquina após desvio de atenção do Governo ao "Kit Covid" [https://saude.ig.com.br/coronavirus/2021-03-28/programa-de-malaria-ficou-sem-estoque-de-cloroquina-apos-desvio-do-governo.html]. |
| Impactos sobre o **prestígio das instituições e do sistema** | Com a desmoralização do Ministério da Saúde (MS) após ter dois Ministros demitidos por não apoiarem o "tratamento precoce" e pelo "vai e vem" retórico [https://www1.folha.uol.com.br/equilibrioesaude/2021/05/apos-reportagem-ministerio-da-saude-apaga-links-com-prescricao-de-cloroquina-para-covid.shtml] diante das evidências cotidianas contra o "tratamento precoce", as próprias práticas do MS baseadas em desinformação desmoralizam sua atuação e a atuação do Estado. Além disso, a disputa com a Agência Nacional de Vigilância Sanitária (ANVISA) e a subida de tom de Bolsonaro ao desrespeitar as próprias normas sanitárias da ANVISA publicamente quase 20 vezes [https://noticias.uol.com.br/cotidiano/ultimas-noticias/reporter-brasil/2021/08/04/bolsonaro-violou-norma-da-anvisa-ao-defender-cloroquina-para-covid.htm] descredibiliza o papel do Estado e passa o recado de que qualquer pessoa pode banaliza-lo. Com a desmoralização do então "tratamento |
| Impactos sobre as **relações domésticas e/ou relações exteriores** | precoce" e pressão da Comissão Parlamentar de Inquérito (CPI) da COVID-19 no Senado, "a cloroquina virou jogo de empurra entre [Ministérios da] Defesa e Saúde", esquivando-se da responsabilidade [https://cnts.org.br/noticias/documentos-mostram-que-cloroquina-virou-jogo-de-empurra-entre-defesa-e-saude-apos-pressao-de-cpi/] e acentuando a desarticulação Federal.<br><br>A narrativa da Hidroxi/Cloroquina impulsionada por Donald Trump, levou Bolsonaro a adotar tal tese "apenas dois dias após Trump falar da cloroquina pela primeira vez" [https://www.nexojornal.com.br/expresso/2021/05/24/Qual-a-cronologia-cient%C3%ADfica-da-cloroquina-na-pandemia]. Mais de um ano após, mesmo com Trump tendo abandonado a tese, o Governo Bolsonaro continuou persistindo na causa. A Organização Mundial da Saúde (OMS) e a União Europeia reforçaram suas preocupações com o Brasil, [https://www.cnnbrasil.com.br/saude/entenda-as-recomendacoes-das-instituicoes-de-saude-contra-o-uso-de-cloroquina/], preocupações estas que se somam às demais que passaram a isolar gradativamente o Brasil nas relações exteriores [https://www.bbc.com/portuguese/internacional-59106310]. |

**Impactos sobre a Estrutura da Sociedade**

| Categoria | Impactos observados |
|---|---|
| Impactos sobre a **adesão à políticas públicas** | Em consequência da empreitada pelo "tratamento precoce" no país, o Governo Bolsonaro conseguiu mobilizar pelo menos 23% da população para o uso de medicações sem comprovações científicas, parcela esta que assumiu ter "usado remédios para tratamento precoce contra a COVID-19" [https://g1.globo.com/ciencia-e-saude/noticia/2021/05/19/datafolha-um-em-cada-quatro-brasileiros-usou-remedios-para-tratamento-precoce-contra-a-covid.ghtml]. Curiosamente, no caso do "tratamento precoce", não se trata tão somente de observar potenciais não-adesões às políticas públicas e às recomendações governamentais, mas também observar a adesão a uma política baseada em desinformação, isto é, a adesão aos medicamentos sem comprovações científicas. |
| Impactos sobre as **alterações na cadeia de produção** | Com o apoio do Governo Federal e insistente interesse na então Hidroxi/Cloroquina [https://politica.estadao.com.br/noticias/geral,documento-de-general-expoe-mapa-da-cloroquina-e-a-cadeia-de-comando-para-produzi-la,70003564204], investigações do Tribunal de Contas da União (TCU) apontam que grupos específicos e farmacêuticas teriam se movimentado antes mesmo da abertura do processo legal de compra. Desta forma, a |
| | desinformação, com apoio na retórica institucional, induziu o setor e levou um custo ao Estado de "167% a mais que compra feita anteriormente [de Cloroquina]" [https://www.cnnbrasil.com.br/nacional/fornecedora-de-cloroquina-do-exercito-foi-consultada-um-mes-antes-de-concorrente/]. O BNDES chegou a estudar a cessão de empréstimos ao setor farmacêutico para produção de Hidroxi/Cloroquina, ação que poderia também ter implicado sobre o estímulo à cadeia de produção [https://aberto.bndes.gov.br/aberto/caso/farmaceuticas/]. Vale citar que internacionalmente setores relevantes no mercado, como a conhecida Kodak, demonstraram interesse em produzir Hidroxi/Cloroquina diante do interesse institucional persistido pelo então presidente Donald Trump [https://www.frontliner.com.br/kodak-ira-fabricar-ingredientes-da-cloroquina/] e, mesmo não sendo no Brasil, reforça a atenção do setor produtivo. |
| Impactos sobre as **alterações na percepção da realidade** | A retórica do "tratamento precoce" e da "imunidade de rebanho" somaram-se às demais teorias da conspiração que circularam a Pandemia da COVID-19: "mais da metade dos brasileiros (56%) acredita na teoria da conspiração que hospitais são pagos para inflar dados de pacientes mortos pela COVID-19"; "44,3% das pessoas que vivem no país acreditam que há uma 'conspiração de esquerda' que quer 'tomar o poder' no Brasil"; "50,7% dos brasileiros considera que o governo da China criou o coronavírus" [https://www.otempo.com.br/brasil/covid-56-dos-brasileiros-creem-em-teoria-da-conspiracao-que-dados-sao-inflados-1.2483512]. Neste sentido, explicita-se a distorção da percepção da realidade pelos brasileiros, acostumando-se cotidianamente com inúmeras notícias falsas, com dificuldades de distinguir estas da realidade. |
| Impactos sobre as **alterações no consumo** | Houve um impacto direto e explícito na demanda, uma vez que esta foi estimulada por meio do "Kit Covid" do Governo Federal, levando a uma alta de 358% no consumo privado do medicamento [https://politica.estadao.com.br/noticias/geral,quem-sao-os-empresarios-que-ganham-com-a-cloroquina-no-brasil,70003360677]. |
| Impactos sobre **externalidades, integridade e ambiente** | Ao final de 2020 foi notificado o crescimento de 558% nos registros anuais de efeitos adversos decorrentes de Hidroxi/Cloroquina, ocasionando inclusive mortes diante do uso desinformado. Dentre os efeitos colaterais observados estão os distúrbios dos sistemas nervoso, gastrointestinal, psiquiátrico e cardíaco [https://www.em.com.br/app/noticia/bem-viver/2021/04/05/interna_bem_viver,1253798/efeitos-adversos-a-cloroquina-disparam-558-e-mortes-sao-registradas.shtml]. |
| Impactos sobre a **opinião pública e sistema de crenças** | Por se tratar do consumo de um medicamento sem previsão de eficácia científica, a alta de 358% no consumo privado do medicamento [https://politica.estadao.com.br/noticias/geral,quem-sao-os-empresarios-que-ganham-com-a-cloroquina-no-brasil,70003360677] e o fato de ao menos 23% da população ter consumido tal medicamento sem comprovações científicas, diante da propagação de desinformação [https://g1.globo.com/ciencia-e-saude/noticia/2021/05/19/datafolha-um-em-cada-quatro-brasileiros-usou-remedios-para-tratamento-precoce-contra-a-covid.ghtml]. Como apontado, contextos de maior vulnerabilidade social absorvem a desinformação com maior incidência [https://www.scielo.br/j/ciedu/a/bW5YKH7YdQ5yZwkJY5LjTts/?lang=pt], reforçando sistemas de crenças especialmente para tal camada social. |
| Impactos sobre o **prestígio do conhecimento científico** | Em um caso explícito de ataque à Ciência, o médico infectologista e cientista pesquisador, Marcus Vinícius de Lacerda, "precisou de escolta por provar que a cloroquina não funciona", além disso, os pesquisadores associados aos seus estudos "começaram a enfrentar uma onda de linchamento nas redes sociais, com ameaças e ataques pessoais" [http://informe.ensp.fiocruz.br/noticias/51215]. Com a crescente das teorias da conspiração em torno da Pandemia da COVID-19 e o apoio institucional do Governo Federal que levou à adesão da população ao "tratamento precoce", notadamente a Ciência e o método científico foi atacado cotidianamente - implicando-se, desta forma, ciclicamente sobre as próprias políticas públicas, estas distanciadas de evidências e legitimadas por retóricas que misturam Ciência com "ponto de vista". |

Fonte: Elaboração própria (2024).

## 5. Referências

## 6. Biografia dos autores

**Ergon Cugler de Moraes Silva** possui mestrado em Administração Pública e Governo (FGV), MBA pós-graduação em Ciência de Dados e Análise (USP) e bacharelado em Gestão de Políticas Públicas (USP). Ele está associado ao Núcleo de Estudos da Burocracia (NEB FGV), colabora com o Observatório Interdisciplinar de Políticas Públicas (OIPP USP), com o Grupo de Estudos em Tecnologia e Inovações na Gestão Pública (GETIP USP), com o Monitor de Debate Político no Meio Digital (Monitor USP) e com o Grupo de Trabalho sobre Estratégia, Dados e Soberania do Grupo de Estudo e Pesquisa sobre Segurança Internacional do Instituto de Relações Internacionais da Universidade de Brasília (GEPSI UnB). É também pesquisador no Instituto Brasileiro de Informação em Ciência e Tecnologia (IBICT), onde trabalha para o Governo Federal em estratégias contra a desinformação. São Paulo, São Paulo, Brasil. Site: https://ergoncugler.com/.

**José Carlos Vaz** é Professor da Universidade de São Paulo - Escola de Artes, Ciências e Humanidades, nos cursos de graduação e de pós-graduação em Gestão de Políticas Públicas. Vice-presidente do Conselho Administrativo do Instituto Pólis. Coordenador do GETIP - Grupo de Estudos em Tecnologia e Inovação na Gestão Pública. Graduação em Administração pela Universidade de São Paulo (1986), Mestrado em Administração Pública pela Fundação Getúlio Vargas SP (1995) e doutorado em Administração de Empresas - Sistemas de Informação pela Fundação Getúlio Vargas SP (2003). Tem experiência na área de Administração Pública, atuando principalmente nos seguintes temas: aspectos sociais e políticos do uso da tecnologia de informação (participação digital, dados governamentais abertos, controle social dos governos, governo eletrônico, políticas públicas de tecnologia), gestão pública (capacidades estatais e de governo, inovações em gestão pública, logística, planejamento estratégico) e questões urbanas e municipais (processos e dinâmicas urbanas, mobilidade urbana, gestão municipal, desenvolvimento local).

-

**Obs:** Levantamento inicialmente realizado por meio do Trabalho de Conclusão de Curso com o título "Como a desinformação impacta políticas públicas?: Elementos para a construção de um modelo analítico para observar potenciais impactos da desinformação e fake news em políticas públicas" (jan. 2022), do curso de Gestão de Políticas Públicas da Escola de Artes, Ciências e Humanidades (EACH), da Universidade de São Paulo (USP), defendido por Ergon Cugler de Moraes Silva, com orientação do professor doutor José Carlos Vaz, contando com a professora doutora Cristiane Kerches da Silva Leite e com o professor doutor Márcio Moretto Ribeiro na banca de defesa. Além disso, foi elaborada a cartilha institucional "Como a desinformação impacta políticas públicas" (set. 2022), com apoio do Grupo de Estudos em Tecnologia e Inovações na Gestão Pública (GETIP) e do Observatório Interdisciplinar de Políticas Públicas "Professor Doutor José Renato de Campos Araújo" (OIPP).